\documentclass[aps,prb,showpacs,eqsecnum,twocolumn,superscriptaddress,amsmath,amssymb]{revtex4-1}

\usepackage[utf8]{inputenc}

\usepackage{graphicx}
\usepackage[colorlinks=true,citecolor=blue]{hyperref}
\usepackage{hypernat}
\usepackage{units}

\usepackage{verbatim}

\newcommand{\bra}[1]{\langle #1|}
\newcommand{\ket}[1]{|#1\rangle}
\renewcommand{\vec}[1]{\mathbf{#1}}
\DeclareMathOperator{\re}{\text{Re}}

\def\up{\uparrow}
\def\down{\downarrow}
\def\kB {k_\text{B}}

\begin{document}

\title{Probing the exchange field of a quantum-dot spin valve by a superconducting lead}

\author{Bj\"orn Sothmann}
\author{David Futterer}
\affiliation{Theoretische Physik, Universit\"at Duisburg-Essen and CeNIDE, 47048 Duisburg, Germany}
\author{Michele Governale}
\affiliation{School of Chemical and Physical Sciences, Victoria University of Wellington, PO Box 600, Wellington 6140, New Zealand}
\author{J\"urgen K\"onig}
\affiliation{Theoretische Physik, Universit\"at Duisburg-Essen and CeNIDE, 47048 Duisburg, Germany}

\date{\today}

\begin{abstract}
Electrons in a quantum-dot spin valve, consisting of a single-level quantum dot coupled to two ferromagnetic leads with magnetizations pointing in arbitrary directions, experience an exchange field that is induced on the dot by the interplay of Coulomb interaction and quantum fluctuations. 
We show that a third, superconducting lead with large superconducting gap attached to the dot probes this exchange field very sensitively.
In particular, we find striking signatures of the exchange field in the symmetric component of the supercurrent with respect to the bias voltage applied between the ferromagnets already for small values of the ferromagnets' spin polarization.
\end{abstract}



\maketitle

\section{Introduction}
The rapidly evolving field of spintronics~\cite{maekawa_spin_2002,maekawa_concepts_2006} not only takes into account the charge degree of freedom of electrons as does conventional electronics, but additionally makes use of the spin degree of freedom thereby opening up interesting new applications as well as addressing questions of fundamental research.
Quantum dots coupled to ferromagnetic electrodes provide one particular example of spintronic systems. They can be experimentally realized in a number of ways, e.g., using metallic nanoparticles,~\cite{deshmukh_using_2002,bernand-mantel_evidence_2006,mitani_current-induced_2008,bernand-mantel_anisotropic_2009} semiconductor quantum dots,~\cite{hamaya_spin_2007,hamaya_electric-field_2007,hamaya_kondo_2007,hamaya_oscillatory_2008,hamaya_tunneling_2008} quantum dots defined in nanowires,~\cite{hofstetter_ferromagnetic_2010} carbon nanotubes,~\cite{jensen_single-wall_2003,sahoo_electric_2005,jensen_hybrid_2004,hauptmann_electric-field-controlled_2008,merchant_current_2009} or single molecules.~\cite{pasupathy_kondo_2004}

From the theoretical point of view, many different properties of such systems have been studied, e.g.,
spin-diode behavior in quantum dots coupled to one ferromagnetic and one normal lead,~\cite{souza_quantum_2007,weymann_spin_2008} the angular dependence of conductance,~\cite{knig_interaction-driven_2003,braun_theory_2004,fransson_angular_2005-1,pedersen_noncollinear_2005,fransson_angular_2005,weymann_cotunneling_2005,weymann_cotunneling_2007,koller_spin_2007} negative tunnel magnetoresistance,~\cite{cottet_magnetoresistance_2006,trocha_quantum_2007,hornberger_transport_2008,trocha_negative_2009} and current noise.~\cite{buka_current_2000,weymann_theory_2007,weymann_shot_2008,souza_spin-polarized_2008,lindebaum_spin-induced_2009} Of particular interest are quantum dots coupled to noncollinearly magnetized ferromagnets.
These systems on the one hand show a nonequilibrium spin accumulation on the quantum dot that has the tendency to block transport.
On the other hand, there is an effective exchange field acting on the dot spin that is caused by virtual tunneling between the quantum dot and the leads and gives rise to a precession of the spin accumulated on the dot.\cite{knig_interaction-driven_2003,braun_theory_2004}
The interplay between these two effects gives rise to a number of interesting transport properties as, e.g., a deviation from the harmonic dependence of the linear conductance on the angle enclosed by the magnetizations,~\cite{knig_interaction-driven_2003} a u-shaped conductance curve with a broad region of negative differential conductance,~\cite{braun_theory_2004} a nontrivial bias dependence of the Fano factor and characteristic features in the finite-frequency noise at the Larmor frequency associated with the exchange field~\cite{braun_frequency-dependent_2006} as well as to a splitting of the Kondo resonance.~\cite{martinek_kondo_2003,martinek_kondo_2003-1,martinek_gate-controlled_2005,utsumi_nonequilibrium_2005,sindel_kondo_2007}
Detecting the exchange field in experiments 
is quite challenging as most of the effects listed above rely on a strong spin blockade of the quantum dot that exists only for large polarizations of the leads. For this reason, the exchange field has so far been detected only by
the induced
splitting of the Kondo resonance in C$_{60}$ molecules,~\cite{pasupathy_kondo_2004} InAs quantum dots,~\cite{hamaya_kondo_2007,hofstetter_ferromagnetic_2010} and carbon nanotubes~\cite{hauptmann_electric-field-controlled_2008} coupled to ferromagnetic leads, respectively.

\begin{figure}
	\includegraphics[width=\columnwidth]{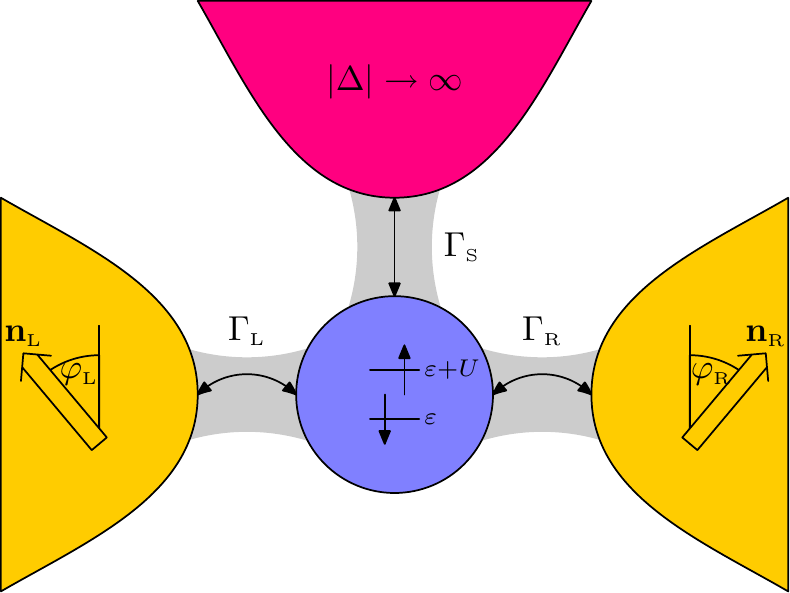}
	\caption{\label{fig:model}(Color online) Schematic model of a quantum-dot spin valve with an additional superconducting electrode attached. A single-level quantum dot with excitation energies $\varepsilon$ and $\varepsilon+U$ is coupled to to noncollinearly magnetized ferromagnets and a superconductor via tunnel barriers.}
\end{figure}

Here, we propose an alternative way to experimentally access the influence of the exchange fields on the transport properties in the
regime of weak tunnel coupling by adding a
superconducting electrode to the quantum dot as shown schematically in Fig.~\ref{fig:model}. Quantum dots coupled to superconducting electrodes are interesting
on their own
as they show an interplay between strong Coulomb interaction that has the tendency to destroy superconducting correlations on the dot and nonequilibrium effects that can help to induce a superconducting proximity effect on the dot. In the subgap regime, transport between the quantum dot and superconductor takes place via Andreev reflections which have been investigated theoretically extensively.~\cite{fazio_resonant_1998,fazio_erratum:_1999,kang_kondo_1998,schwab_andreev_1999,clerk_andreev_2000,cuevas_kondo_2001,pala_nonequilibrium_2007,governale_real-time_2008,futterer_nonlocal_2009}
 Further studies involved multiple Andreev reflections~\cite{yeyati_resonant_1997,johansson_resonant_1999} and transport in the Kondo regime.~\cite{clerk_loss_2000,avishai_superconductor-quantum_2003,sellier_pi_2005,nussinov_spin_2005,bergeret_interplay_2006,lpez_josephson_2007,karrasch_josephson_2008}
Experimentally, quantum dots coupled to superconductors have been realized using carbon nanotubes,~\cite{buitelaar_quantum_2002,cleuziou_carbon_2006,jarillo-herrero_quantum_2006,jrgensen_electron_2006,jrgensen_critical_2007,dirks_superconducting_2009,herrmann_carbon_2010} graphene,~\cite{dirks_andreev_2010} semiconductor nanowires,~\cite{van_dam_supercurrent_2006,sand-jespersen_kondo-enhanced_2007,hofstetter_ferromagnetic_2010} self-assembled semiconductor quantum dots,~\cite{buizert_kondo_2007} and single molecules.~\cite{winkelmann_superconductivity_2009}

For the system under investigation, we compute the current into the superconductor using a real-time diagrammatic approach~\cite{knig_zero-bias_1996,knig_resonant_1996,schoeller_transport_1997,knig_quantum_1999,braun_theory_2004,pala_nonequilibrium_2007,governale_real-time_2008,futterer_nonlocal_2009} in the limit of an infinite superconducting gap. This is a reasonable approximation as long as the excitation energies of the quantum dot are smaller than the gap such that sub-gap transport takes place. The current has, in general,  even and  odd components with respect to the voltage applied between the ferromagnets. We find that the
even component exhibits clear evidences  of the exchange field even for small polarizations of the ferromagnets. 

The paper is organized as follows. We introduce the model for the quantum dot coupled to the three electrodes in Sec.~\ref{sec:model}.
In Sec.~\ref{sec:technique}, we describe how to apply the real-time diagrammatic theory to the system under investigation. We present our results in Sec.~\ref{sec:results} and give a conclusion in Sec.~\ref{sec:conclusions}.

\section{\label{sec:model}Model}

We consider a quantum-dot spin valve, i.e., a quantum dot coupled to two ferromagnetic electrodes with magnetizations pointing in arbitrary directions $\vec n_\text{L}$, $\vec n_\text{R}$.
In addition to the two ferromagnetic electrodes we consider a third superconducting lead
tunnel coupled to the quantum dot. 
The Hamiltonian of the system hence consists of different terms 
describing the ferromagnetic electrodes, the superconducting electrode, the quantum dot and the tunnel coupling between the dot and the leads,
\begin{equation}
	H=\sum_rH_r+H_\text{S}+H_\text{dot}+H_\text{tun}.
\end{equation}
We describe the  ferromagnetic electrodes with chemical potentials $\mu_r$, $r=\text{L},\text{R}$, by means of the free-electron Hamiltonians 
\begin{equation}
	H_r=\sum_{\vec k\sigma}(\varepsilon_{r\vec k\sigma}-\mu_r)a_{r\vec k\sigma}^\dagger a_{r\vec k\sigma},
\end{equation}
where the quantization axis is chosen to be parallel to the magnetization of the respective lead. We assume the densities of states $\rho_{r\sigma}(\omega)=\sum_\vec k\delta(\omega-\varepsilon_{r\vec k\sigma})$ to be constant, $\rho_{r\sigma}(\omega)=\rho_{r\sigma}$, and spin dependent. The asymmetry between majority ($\sigma=+$) and minority ($\sigma=-$) spins is  characterized by the polarization $p_r=(\rho_{r+}-\rho_{r-})/(\rho_{r+}+\rho_{r-})$ which varies between $p=0$ for a nonmagnetic electrode and $p=1$ for a halfmetallic electrode with majority spins only.

We model the superconductor by means of a mean-field BCS Hamiltonian with a superconducting gap $\Delta$, which can be chosen to be real and positive without loss of generality.
 We choose the chemical potential of the superconductor as  reference for energies and set it to zero.
In the limit of an infinite superconducting gap, $\Delta\to\infty$, the single-level quantum dot tunnel coupled to the superconductor is
described   by the effective dot Hamiltonian~\cite{rozhkov_interacting-impurity_2000,karrasch_supercurrent_2009,meng_self-consistent_2009}
\begin{equation}\label{eq:HDotEff}
	H_\text{dot,eff}=\sum_{\sigma}\varepsilon c_{\sigma}^\dagger c_{\sigma}+U c_{\up}^\dagger c_{\up}c_{\down}^\dagger c_{\down}-\frac{\Gamma_\text{S}}{2}c_\up^\dagger c_\down^\dagger-\frac{\Gamma_\text{S}}{2}c_\down c_\up,
\end{equation}
where  $\varepsilon$ is the energy of the spin-degenerate level in the dot and
$U$ denotes the Coulomb energy for double occupancy of the dot.
The effective pair potential  $\Gamma_\text{S}$ in Eq.~(\ref{eq:HDotEff}) is the tunnel-coupling strength between the dot and the superconductor and it is related to microscopic parameters by $\Gamma_{\text{S}}=2\pi |t_{\text{S}}|^2 \rho_{\text{S}}$, where $t_{\text{S}}$ is the tunneling amplitude between dot and superconductor  and $ \rho_{\text{S}}$ is the normal-state density of states of the superconducting lead.

The effective dot Hamiltonian accounts for the coupling to the superconductor exactly. This allows us to deal with an arbitrarily strong superconductor-dot coupling $\Gamma_\text{S}$. The eigenstates of the effective dot Hamiltonian~\eqref{eq:HDotEff} are given by the singly occupied states $\ket{\up}$ and $\ket{\down}$ as well as by the two states $\ket{+}$ and $\ket{-}$. The latter ones are linear combinations of the empty and doubly occupied dot states $\ket{0}$ and $\ket{d}=c^\dagger_\uparrow c^\dagger_\downarrow |0\rangle $:
\begin{align}
	\ket{+}&=\frac{1}{\sqrt{2}}\left(\sqrt{1-\frac{\delta}{2\varepsilon_\text{A}}}\ket{0}-\sqrt{1+\frac{\delta}{2\varepsilon_\text{A}}}\ket{d}\right),\\
	\ket{-}&=\frac{1}{\sqrt{2}}\left(\sqrt{1+\frac{\delta}{2\varepsilon_\text{A}}}\ket{0}+\sqrt{1-\frac{\delta}{2\varepsilon_\text{A}}}\ket{d}\right).
\end{align}
The energies of the eigenstates are given by $E_\up=E_\down=\varepsilon$ and $E_\pm=\delta/2\pm \varepsilon_\text{A}$. Here, $\delta=2\varepsilon+U$ denotes the detuning from the particle-hole symmetry point while $2\varepsilon_\text{A}=\sqrt{\delta^2+\Gamma_\text{S}^2}$ measures the energy difference between the states $\ket{+}$ and $\ket{-}$. 

We define the  Andreev bound-state energies as the excitation energies of the dot in the absence of tunnel coupling to the ferromagnets:
\begin{equation}
	E_{\text{A},\gamma'\gamma}=\gamma'\frac{U}{2}+\frac{\gamma}{2}\sqrt{\delta^2+\Gamma_\text{S}^2},\quad \gamma,\gamma'=\pm.
\end{equation}

As we allow arbitrarily oriented magnetizations of the ferromagnetic leads, $\vec n_\text{L}$, $\vec n_\text{R}$, it turns out to be most convenient to quantize the spin on the quantum dot in the direction of $\vec n_\text{L}\times\vec n_\text{R}$. In this case, the tunnel coupling between the dot and the ferromagnets is characterized by
\begin{multline}
	H_\text{tun,F}=\sum_{r\vec k}\frac{t_r}{\sqrt{2}}\left[a_{r\vec k+}^\dagger\left(e^{i\phi_r}c_\up+e^{-i\phi_r}c_\down\right)\right.\\\left.
	+a_{r\vec k-}^\dagger\left(-e^{i\phi_r}c_\up+e^{-i\phi_r}c_\down\right)\right]+\text{h.c.},
\end{multline}
i.e., the majority/minority spin electrons of the leads couple to both spin up and spin down states of the quantum dot. In the tunnel Hamiltonian, $\phi_\text{L}=-\phi_\text{R}=\phi/2$ is half the angle between the magnetizations. The tunnel matrix elements $t_r$ can be related to the spin-dependent tunnel couplings $2\pi|t_r|^2\rho_{r\sigma}$. The total tunnel coupling to lead $r$ is given by $\Gamma_r=\sum_{\sigma}2\pi|t_r|^2\rho_{r\sigma}/2$.

\section{\label{sec:technique}Technique}
In order to compute the transport properties of the quantum-dot spin valve with an additional superconducting lead, we use the diagrammatic real-time technique~\cite{knig_zero-bias_1996,knig_resonant_1996,schoeller_transport_1997,knig_quantum_1999} in its extension to ferromagnetic~\cite{braun_theory_2004} and superconducting~\cite{pala_nonequilibrium_2007,governale_real-time_2008,futterer_nonlocal_2009} leads.

The basic idea of this approach is to integrate out the noninteracting, fermionic degrees of freedom of the electrodes. The remaining system consisting of the quantum dot is then described using a reduced density matrix $\rho^\text{red}$ with matrix elements $P_{\chi_1}^{\chi_2}=\bra{\chi_2}\rho^\text{red}\ket{\chi_1}$. For convenience, we introduce for the diagonal matrix elements the abbreviation $P_\chi=P_\chi^\chi$.

For a quantum dot coupled to both ferromagnetic and superconducting leads the reduced density matrix in the stationary limit takes the form
\begin{equation}
	\rho^\text{red}=
	\left(
	\begin{array}{cccc}
		P_+ & 0 & 0 & 0 \\
		0 & P_- & 0 & 0 \\
		0 & 0 & P_\up & P^\up_\down \\
		0 & 0 & P^\down_\up & P_\down
	\end{array}
	\right)
\end{equation}
in the basis $\ket{+}$, $\ket{-}$, $\ket{\up}$, and $\ket{\down}$. We note that while coherent superpositions of states $\ket{\up}$ and $\ket{\down}$ have to be taken into account as we are dealing with a noncollinear geometry, coherent superpositions of $\ket{+}$ and $\ket{-}$ can be neglected if $\Gamma_\text{S}\gg\Gamma_r$ (which we assume from now on) as they are at least of order $O(\Gamma_r)$ and therefore do not contribute to transport in
the limit of weak tunnel coupling between dot and ferromagnets.

In the stationary state, the reduced density matrix obeys a master equation of the form
\begin{equation}\label{eq:master}
	0=\dot P_{\chi_1}^{\chi_2}=-i(E_{\chi_2}-E_{\chi_1})P_{\chi_1}^{\chi_2}+\sum_{\chi_1'\chi_2'}W_{\chi_1\chi_1'}^{\chi_2\chi_2'}P_{\chi_1'}^{\chi_2'}.
\end{equation}
While the first term on the right-hand side describes the coherent evolution of the system, the second term characterizes the dissipative coupling to the electrodes. The quantities $W_{\chi_1\chi_1'}^{\chi_2\chi_2'}$ are generalized transition rates in Liouville space. For tunnel couplings small compared to the temperature, $\Gamma_r\ll\kB T$, they can be evaluated in a perturbation expansion in the tunnel couplings as irreducible self-energy blocks of the dot propagator on the Keldysh contour. The corresponding diagrammatic rules are summarized in Appendix~\ref{sec:AppRules}.

Using the effective dot description and introducing the average spin on the dot as
\begin{equation}
	S_x=\frac{P_\up^\down+P^\up_\down}{2}, S_y=\frac{P_\up^\down-P^\up_\down}{2i}, S_z=\frac{P_\up-P_\down}{2},
\end{equation}
as well as the probability $P_1=P_\up+P_\down$ to find the dot singly occupied, we can split the master equation into one set for the occupation probabilities and one set for the average spin. We introduce the abbreviations: $\Gamma_{r\pm}=\Gamma_r\left(1\pm\frac{\delta}{2\varepsilon_\text{A}}\right)$ and $f_{r\gamma\gamma'}^\pm=f_r^\pm(E_{\text{A},\gamma\gamma'})$, where $f_r^+(\omega)=1-f_r^-(\omega)$ denotes the Fermi function of lead $r$. 
The equations governing the occupation probabilities are given by
\begin{equation}
\label{eq-diagonal}
	0=\frac{d}{dt}
	\left(
	\begin{array}{c}
		P_+ \\
		P_- \\
		P_1
	\end{array}
	\right)
	=
	\sum_r \mathbf{A}_r
 	\left(
	\begin{array}{c}
		P_+ \\
		P_- \\
		P_1
	\end{array}
	\right)\\
	+
	\sum_r
	p_r \mathbf{b}_r
	(\vec S_r\cdot \vec n_r), 
\end{equation}
where the expressions for the  matrices $\mathbf{A}_r$ and the vectors  $\mathbf{b}_r$ are given in Appendix \ref{appendixb}.
We see that similar to the case of an ordinary quantum-dot spin valve,~\cite{braun_theory_2004} the dynamics of the occupation probabilities couples to the average spin accumulated on the quantum dot.

\begin{figure}
	\includegraphics[width=\columnwidth]{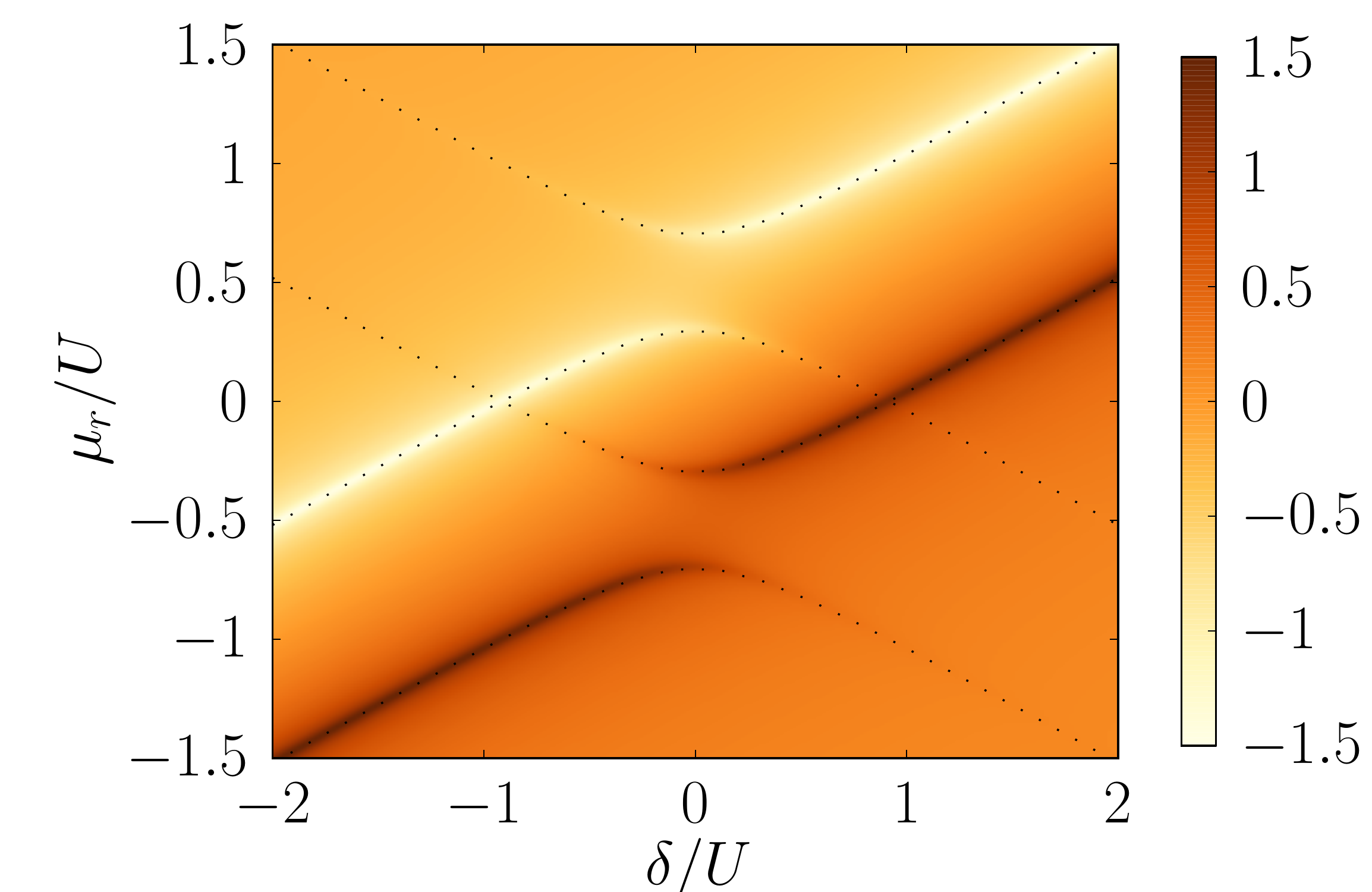}
	\caption{\label{fig:exchangefield}(Color online)
	Amplitude of the exchange field $\vec B_r\cdot \vec{n}_r$ in units of $p_r\Gamma_r$ for $  k_{\rm B} T=0.01U$, $\Gamma_\text{S}=0.4U$ as a function of the chemical potential $\mu_r$ and detuning $\delta$. The peaks and dips map out the Andreev bound states whose energies are indicated by dotted lines.}
\end{figure}

As in the case of the normal quantum-dot spin valve, the master equation for the average dot spin can be cast into the form of a Bloch equation,
\begin{equation}\label{eq:masterSpin}
	\frac{d\vec S}{dt}=\left(\frac{d\vec S}{dt}\right)_\text{acc}+\left(\frac{d\vec S}{dt}\right)_\text{rel}+\left(\frac{d\vec S}{dt}\right)_\text{prec},
\end{equation}
where the first term
\begin{multline}
	\left(\frac{d\vec S}{dt}\right)_\text{acc}=\frac{1}{2}\sum_r \left[\left(-\Gamma_{r+}f_{r++}^-+\Gamma_{r-}f_{r--}^+\right)P_+\right.\\\left.+\left(-\Gamma_{r-}f_{r+-}^-+\Gamma_{r+}f_{r-+}^+\right)P_-\phantom{\frac{1}{2}}\right.\\\left.
	+\frac{1}{2}\left(-\Gamma_{r-}f_{r--}^--\Gamma_{r+}f_{r-+}^-\right.\right.\\\left.\left.+\Gamma_{r+}f_{r++}^++\Gamma_{r-}f_{r+-}^+\right)P_1\right] p_r \vec n_r,
\end{multline}
describes the nonequilibrium spin accumulation on the dot due to spin-dependent tunneling of electrons onto the dot. The relaxation of the dot spin is described by the second term,
\begin{multline}
	\left(\frac{d\vec S}{dt}\right)_\text{rel}=-\frac{1}{2}\sum_r\left(\Gamma_{r-}f_{r--}^-+\Gamma_{r+}f_{r-+}^-\right.\\\left.+\Gamma_{r+}f_{r++}^++\Gamma_{r-}f_{r+-}^+\right)\vec S,
\end{multline}
which is proportional to the spin accumulated on the dot. The dot spin relaxes either by electrons with a given spin leaving the dot to the ferromagnetic leads or by electrons with a spin opposite to that on the dot entering the dot from the ferromagnets, thus forming a spin singlet. Finally, the last term
\begin{equation}
	\left(\frac{d\vec S}{dt}\right)_\text{prec}=\sum_r \vec S\times\vec B_r,
\end{equation}
describes a precession of the dot spin due to an exchange field which is given by
\begin{equation}
	\vec B_r=\frac{p_r\vec n_r}{2\pi}\sum_{\gamma \gamma'} \gamma' \Gamma_{r\gamma}
	\re\Psi\left(\frac{1}{2}+i\frac{\beta (E_{\text{A},\gamma'\gamma}-\mu_r)}{2\pi}\right),
\end{equation}
where $\Psi$ is the digamma function. The exchange field is the manifestation of a spin-dependent level renormalization due to virtual tunneling between the dot and the ferromagnetic electrodes. We emphasize that the coupling to the superconductor influences the exchange field only through the position of the Andreev bound states. As can be seen in Fig.~\ref{fig:exchangefield}, the exchange field takes on large values whenever one of the Andreev bound states is in resonance with the Fermi level of the ferromagnet. This behavior is similar to the ordinary quantum-dot spin valve where the exchange field becomes maximal at resonance as well.

The particle current flowing from the ferromagnetic leads into the quantum dot is given by
\begin{equation}\label{eq:current}
	I_r=\sum_{\chi\chi_1\chi_2} W^{I_r\chi\chi_2}_{\phantom{I_r}\chi\chi_1} P_{\chi_1}^{\chi_2}.
\end{equation}
Here, $W^{I_r\chi\chi_2}_{\phantom{I_r}\chi\chi_1}$ are the current kernels that can be obtained from the generalized transition rates $W^{\chi \chi_2}_{\chi\chi_1}$ by multiplying the rate with the number of electrons transferred from lead $r$ to the quantum dot in the process associated with this rate.
We find
\begin{eqnarray}\label{eq:currentF}
	I_r&=& \left(\Gamma_{r-}f_{r--}^+-\Gamma_{r+}f_{r++}^-\right)P_+ \nonumber \\
	&&+\left(\Gamma_{r+}f_{r-+}^+-\Gamma_{r-}f_{r+-}^-\right)P_- \nonumber\\
	&&+\frac{1}{2}\left(\Gamma_{r+}f_{r++}^++\Gamma_{r-}f_{r+-}^+
	\right. \nonumber \\ && \left.
	-\Gamma_{r-}f_{r--}^--\Gamma_{r+}f_{r-+}^-\right)P_1 \nonumber \\
 	&&-p_r \left(\Gamma_{r-}f_{r--}^-+\Gamma_{r+}f_{r-+}^- 
	\right. \nonumber \\ && \left.
 	+\Gamma_{r+}f_{r++}^++\Gamma_{r-}f_{r+-}^+\right)\vec S\cdot\vec n_r.
\end{eqnarray}
In the stationary state, the current into the superconductor is related to the currents between the dot and the ferromagnets by current conservation, which is automatically satisifed in the real-time diagrammatic theory,~\cite{knig_zero-bias_1996,knig_resonant_1996,schoeller_transport_1997,knig_quantum_1999}
\begin{equation}\label{eq:currentS}
	I_\text{S}=I_\text{L}+I_\text{R}.
\end{equation}
Experimentally, one can therefore measure either the current flowing into the superconductor or the difference between the currents that enter from the left and leave to the right ferromagnet. In the following discussion we will however always refer to the current into the superconductor for simplicity.

\section{\label{sec:results}Results}
In this section, we discuss how the current into the superconductor can be used to probe the exchange field. 
We consider symmetrically biased ferromagnets, $\mu_\text{L}=-\mu_\text{R}\equiv\mu$, while the superconductor is kept at $\mu_\text{S}=0$.
We  split the current into a component that is a symmetric function of bias, $I_\text{S}^\text{sym}(\mu)=[ I_\text{S}(\mu)+I_\text{S}(-\mu)]/2$, and a component that is an antisymmetric function of bias, $I_\text{S}^\text{antisym}(\mu)=[I_\text{S}(\mu)-I_\text{S}(-\mu)]/2$. The symmetric component of the current turns out to be very sensitive to the exchange field.

\subsection{\label{ssec:spin-controlled}Symmetric quantum-dot spin valve}
\begin{figure}
	\includegraphics[width=.49\columnwidth]{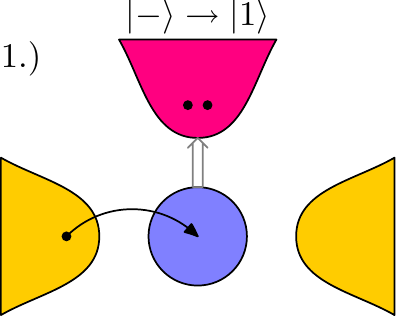}
	\includegraphics[width=.49\columnwidth]{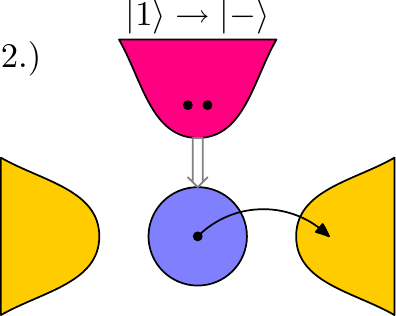}
	\includegraphics[width=.49\columnwidth]{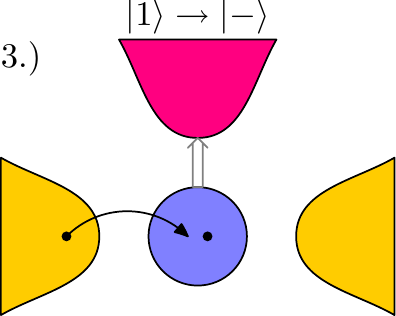}
	\includegraphics[width=.49\columnwidth]{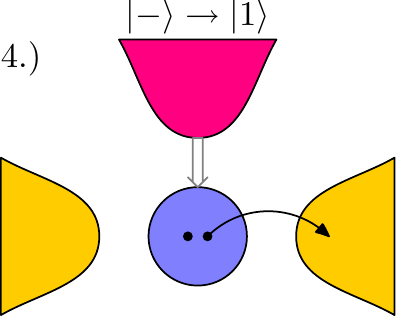}
	\caption{\label{fig:processes}(Color online)
	Transport processes in the proximized quantum-dot spin valve. In the first process, the $\ket{-}$ state is projected onto the $\ket{0}$ state, pushing a Cooper pair into the superconductor as indicated by the double arrow. Then an electron tunnels in from the left lead, leaving the dot in the singly occupied state. Similarly, in the second process, an electron leaves the singly occupied dot to the right lead. Then, a Cooper pair enters from the superconductor to bring the dot in the state $\ket{-}$. As both processes probe the empty contribution to $\ket{-}$, their rates are proportional to $1+\delta/(2\varepsilon_\text{A})$. Similarly, the other two processes probe the doubly occupied component of $\ket{-}$ such that their rates are proportional to $1-\delta/(2\varepsilon_\text{A})$.}
\end{figure}
We start our discussion by considering a symmetric system, i.e., we assume the tunnel couplings to the left and right ferromagnet to be equal, $\Gamma_\text{L}=\Gamma_\text{R}\equiv\Gamma/2$. Furthermore, we assume both ferromagnets to have the same polarization $p$. 
In this particular case, due to symmetry, the current possesses only a symmetric component with respect to $\mu$, i.e.  $I_\text{S}(\mu)=I_\text{S}^\text{sym}(\mu)$.

We first explain why the supercurrent vanishes for collinear geometries. We then show that the spin accumulation in the noncollinear configuration gives rise to a finite current into the superconductor that is sensitive to the exchange field. Finally, we show that a spin relaxation on the dot reduces the supercurrent but nevertheless still allows a detection of the exchange field.

In general, the current into the superconductor vanishes in the small-bias regime $E_{\text{A}-+}<\mu<E_{\text{A}+-}$, where the quantum dot is Coulomb-blockaded. For a symmetric system, the current also vanishes in the large-bias regime, $\mu>E_{\text{A}++}$ or $\mu<E_{\text{A}--}$, where all dot states contribute to transport, due to particle-hole symmetry. Hence, we can expect a finite current into the superconductor only in the intermediate bias regime, $E_{\text{A}--}<\mu<E_{\text{A}-+}$ or $E_{\text{A}+-}<\mu<E_{\text{A}++}$. According to Eq.~\eqref{eq:currentF} and~\eqref{eq:currentS}, the supercurrent in this regime is given by
\begin{multline}\label{eq:currentSIntermediate}
	I_\text{S}=\frac{\Gamma}{2}\left(-2\frac{\delta}{\varepsilon_\text{A}}P_++\frac{\delta}{\varepsilon_\text{A}}P_--\frac{\delta}{2\varepsilon_\text{A}}P_1\right.\\\left.
	-p\vec S\cdot\left(\vec n_\text{L}+\vec n_\text{R}\right)+\frac{p\delta}{2\varepsilon_\text{A}}\vec S\cdot\left(\vec n_\text{L}-\vec n_\text{R}\right)\right).
\end{multline}
For parallel magnetizations, where the spin accumulation on the dot vanishes and the dot occupation probabilities satisfy $P_1=2P_-=1/3$, we find that the supercurrent vanishes also in the intermediate bias regime. To understand the mechanism behind this behavior, let us consider the transport processes that contribute to the superconductor. Notice that in the intermediate regime the state $|+\rangle$ is inaccessible. We find that the contributions from the first two processes shown in Fig.~\ref{fig:processes} cancel each other. They both transfer equal amounts of charge between the dot and the superconductor when projecting the state $\ket{-}$ onto the state $\ket{0}$. Furthermore, both processes have identical rates (a factor of 2 due to spin is compensated by $P_1=2P_-$). In consequence, they give rise to a vanishing supercurrent for any value of the detuning $\delta$. Similarly, one can show that the other two processes that probe the doubly occupied component of $\ket{-}$ cancel each other.

The situation is more complex in the antiparallel configuration due to the finite spin accumulation on the quantum dot. Processes 1 and 4 in Fig.~\ref{fig:processes} which build up the spin accumulation have rates proportional to $1+\delta/(2\varepsilon_\text{A})$ and $1-\delta/(2\varepsilon_\text{A})$, respectively. Hence, in sum the spin accumulation is insensitive to the detuning $\delta$. A similar reasoning holds for processes 2 and 3 that relax the dot spin. As the supercurrent vanishes at $\delta=0$ due to particle-hole symmetry, it therefore has to vanish for all values of $\delta$.

\begin{figure}
	\includegraphics[width=\columnwidth]{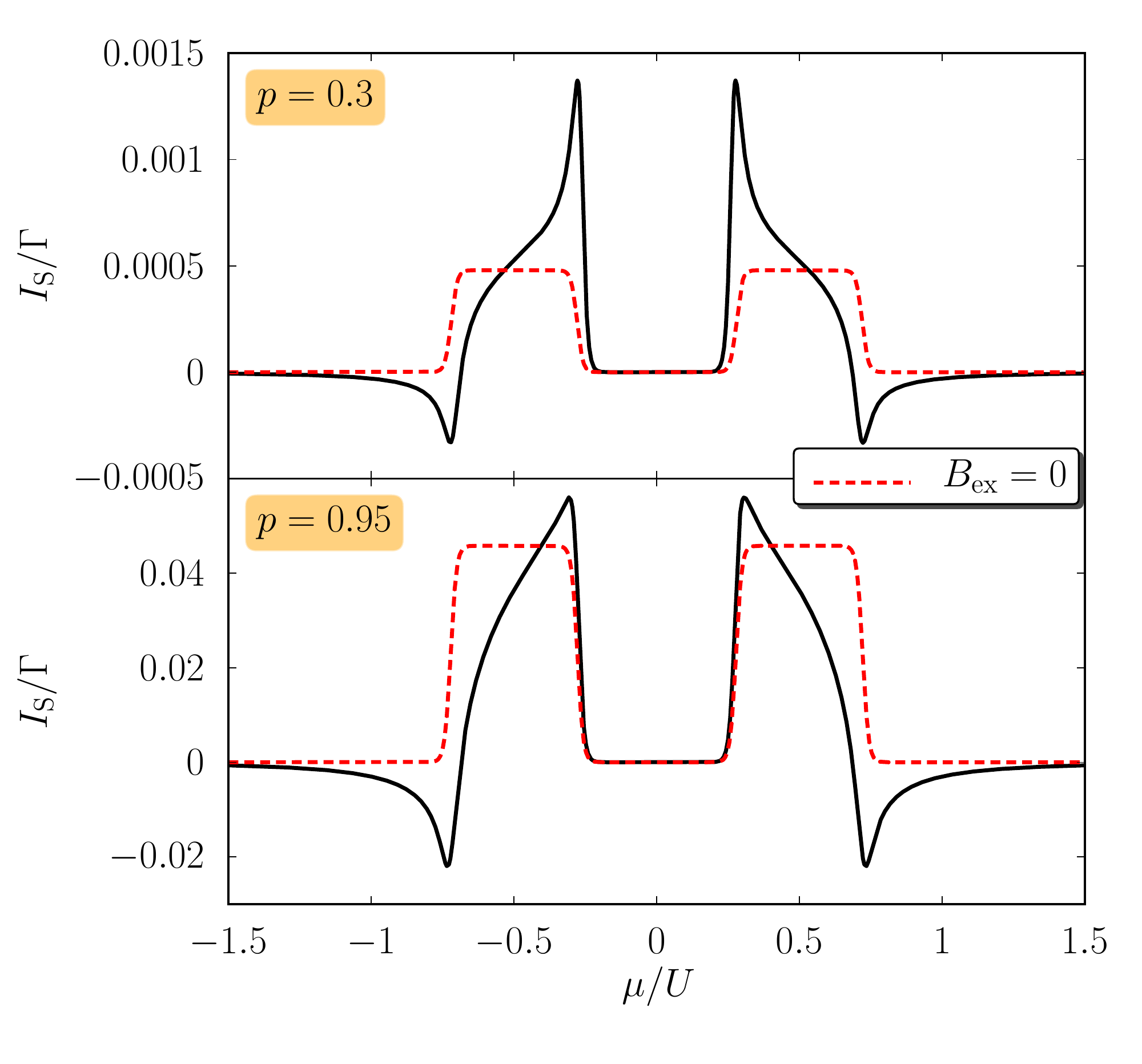}
	\caption{\label{fig:currentS}(Color online) Current into the superconductor for a symmetric coupling, $\Gamma_\text{L}=\Gamma_\text{R}$, $\mu_\text{L}=-\mu_\text{R}\equiv\mu$ and perpendicularly magnetized ferromagnets, $\phi=\pi/2$. In the upper panel, we have $p=0.3$, while in the lower panel $p=0.95$. The solid (black) curves take into account the exchange field, it is neglected in the dashed (red) curves. Other parameters are $\delta=0.2U$, $\Gamma_\text{S}=0.4U$, $  k_{\rm B} T=0.01U$.}
\end{figure}

For noncollinear geometries, we find a finite current into the superconductor, cf. Fig.~\ref{fig:currentS}. To understand the mechanism leading to the finite current, we first neglect the exchange field in our discussion and turn to its effect afterwards.

In contrast to the antiparallel configuration, in the noncollinear geometries the spin accumulation on the quantum dot (and therefore also the probability to find the dot singly occupied) is sensitive to $\delta$. This can be understood by considering again the processes 1 and 4 of Fig.~\ref{fig:processes} which are responsible for the spin accumulation. While process 1 builds up a dot spin in in the direction of $+\vec n_\text{L}$, process 4 build up a spin in the direction of $-\vec n_\text{R}$. For $\delta=0$ both processes contribute equally to the spin accumulation. We therefore find that a spin builds up in the direction of $\vec n_\text{L}-\vec n_\text{R}$. For positive $\delta$, process 1 dominates and hence the spin points toward $\vec n_\text{L}$, while for negative $\delta$, process 4 dominates and the spin points toward $-\vec n_\text{R}$. As for finite $\delta$ the left-right symmetry is broken, the cancellation between the supercurrent contributions from processes 1 and 2 (3 and 4, respectively) does not hold any longer and a finite supercurrent can flow.
Neglecting the exchange field, the following analytic expression for the current into the superconductor can be found:
\begin{equation}\label{eq:ISsym}
	I_\text{S}=\frac{\Gamma\Gamma_\text{S}^2p^4\delta\sin^2\phi}{\varepsilon_\text{A}\left(48\varepsilon_\text{A}^2-2p^2(2\delta^2(1+2\cos\phi)-\Gamma_\text{S}^2(1-\cos\phi))\right)}.
\end{equation}
The above formula shows that the current into the supercurrent in the noncollinear geometry flows for any value of the polarization $p\neq0$. It is only the magnitude of the current that is affected by the strength of the polarization. Using realistic parameters, $p=0.3$, $U\sim\unit[1]{meV}$, $\Gamma_\text{S}\sim\unit[0.5]{meV}$, $\Gamma\sim\unit[100]{\mu eV}$, and $\phi=\pi/2$, we obtain as an order of magnitude of the current $I_\text{S}\sim\unit[1]{pA}$ which is challenging but not impossible to measure with current techniques.

If the exchange field is taken into account, there is still a finite current flowing in the intermediate bias regime. As the dot spin now precesses in the energy-dependent exchange field, it acquires a finite $z$ component while simultaneously the $x$ and $y$ components which influence the supercurrent, cf. Eq.~\eqref{eq:currentSIntermediate}, get reduced and show a nontrivial bias dependence. In consequence, the supercurrent also deviates from its steplike behavior in the absence of the exchange field and even changes sign.
Furthermore, there is a finite supercurrent flowing in the large bias regime because the symmetry-breaking spin accumulation on the dot persists in this regime.

\begin{figure}
	\includegraphics[width=\columnwidth]{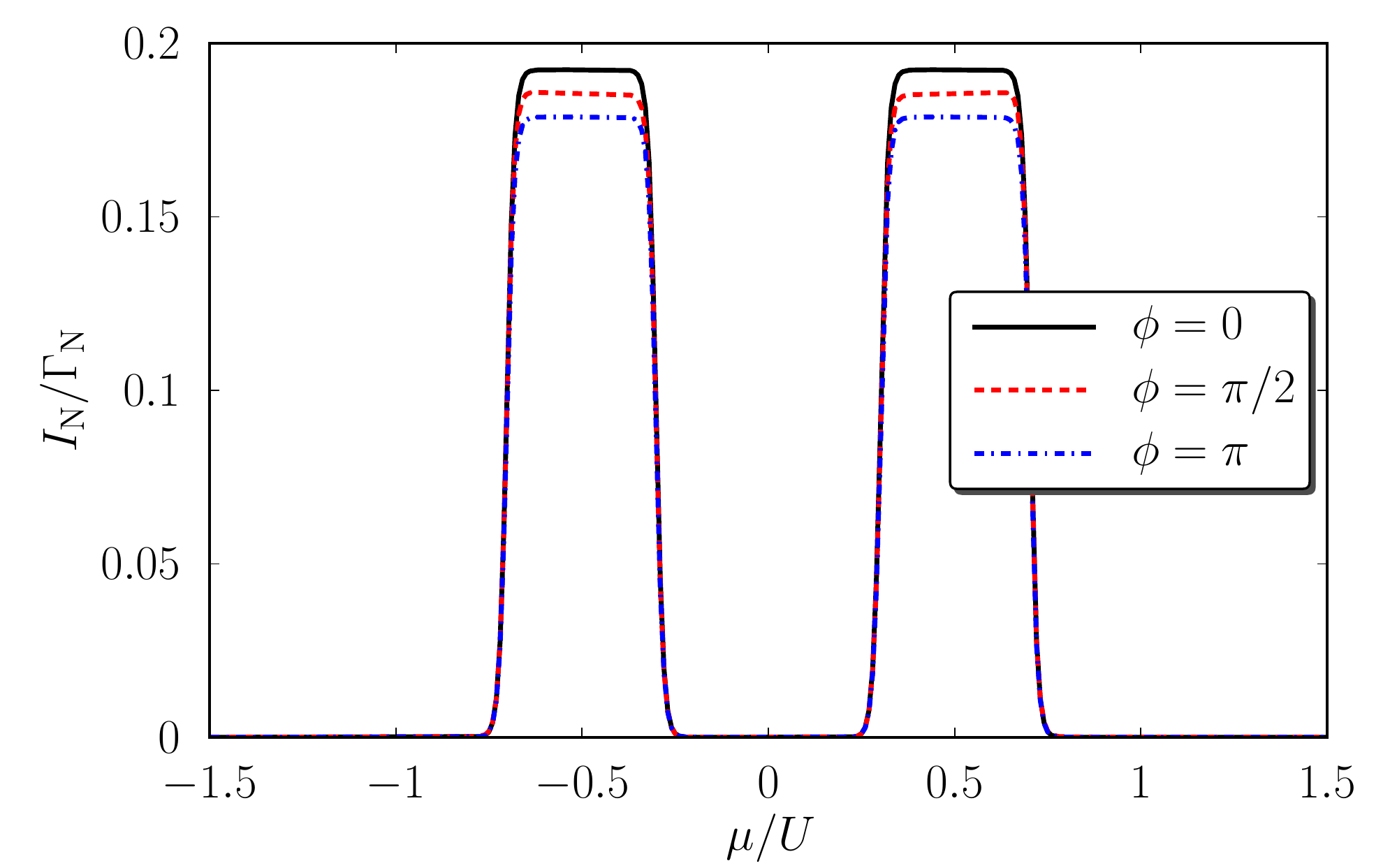}
	\caption{\label{fig:currentFDFN}(Color online) When the third lead is a normal metal instead of a superconductor, a finite current flows for any magnetic configuration, thereby completely obscuring the exchange field effects for small polarizations. Parameters are $\delta=0.4U$, $\Gamma_r=\Gamma_\text{N}$, $p=0.3$, $  k_{\rm B} T=0.01U$.}
\end{figure}

The nontrivial bias dependence of the supercurrent opens up the possibility to detect the exchange field experimentally, even for small polarizations. This is in strong contrast to the other exchange field effects that arise in the sequential tunneling regime, as, e.g., negative differential conductance~\cite{braun_theory_2004,hornberger_transport_2008} or the nontrivial dependence of current, Fano factor and higher current cumulants on the angle between the magnetizations.~\cite{braun_theory_2004,braun_frequency-dependent_2006,lindebaum_spin-induced_2009} While for quantum dots that couple only to ferromagnetic leads all exchange field effects rely on a strong spin blockade, in the system under investigation here, it is the cancellation between different transport processes involving the superconductor that provides the necessary sensitivity to the spin accumulation and the exchange field. To illustrate this, let us consider a system where the superconductor is replaced by a normal metal that is coupled to the quantum dot with coupling strength $\Gamma_\text{N}$. In Fig.~\ref{fig:currentFDFN} we show the current into the normal metal evaluated to first order in $\Gamma_\text{N}$, an approximation valid if $\Gamma_\text{N}\ll\kB T$. We find that now there is indeed a finite current for all magnetic configurations because for a positive (negative) detuning only processes where electrons leave (enter) the dot to (from) the third lead are possible. In consequence, the exchange field effects become practically invisible for small polarizations as they are obfuscated by the large background current.

\begin{figure}
	\includegraphics[width=\columnwidth]{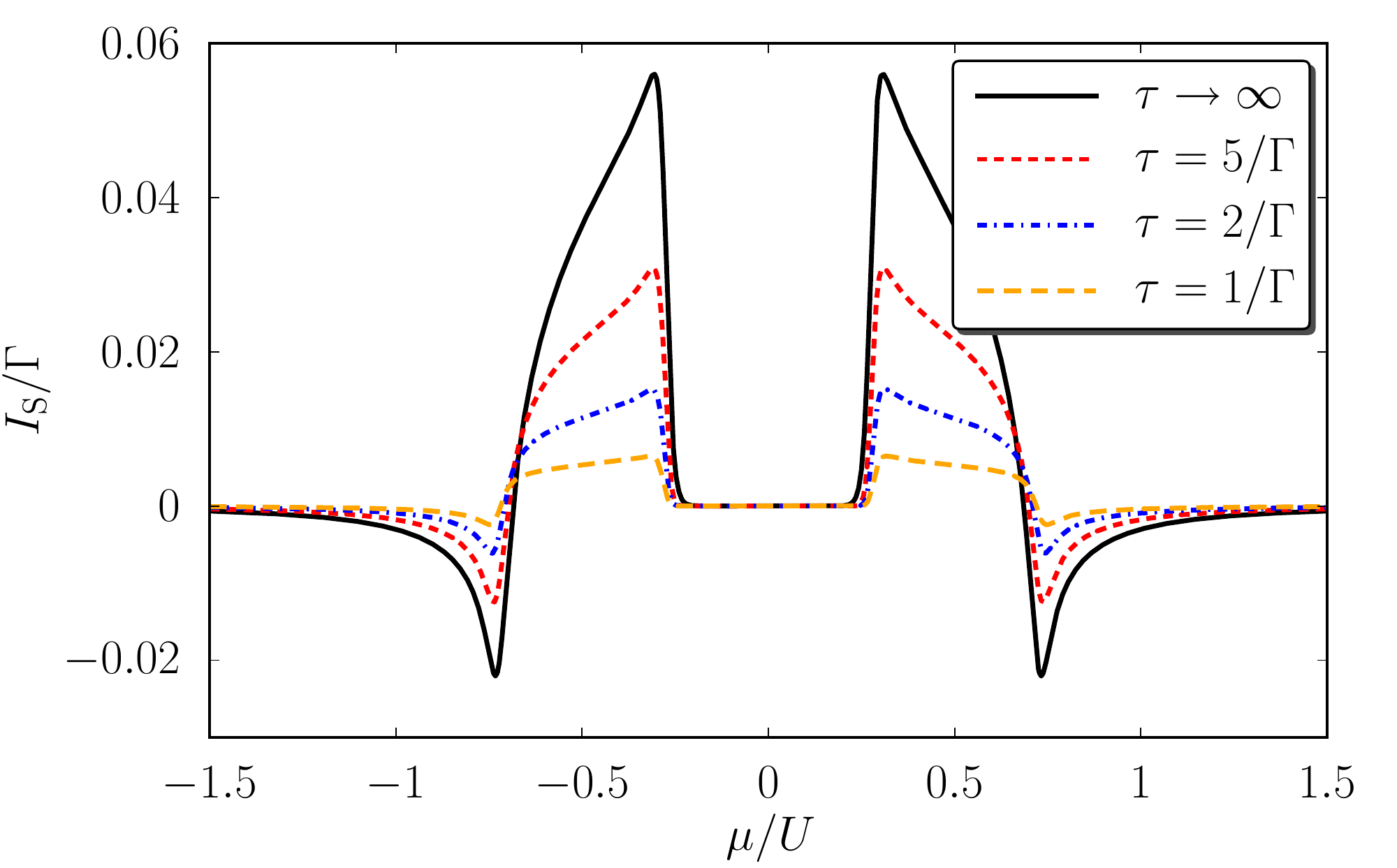}
	\caption{\label{fig:relaxation}(Color online) Influence of spin relaxation on the supercurrent in the noncollinear configuration. As the spin accumulation is reduced, the supercurrent is decreased. However, the exchange field effects still prevail. Polarization is $p=0.95$, other parameters as in Fig.~\ref{fig:processes}.}
\end{figure}

Finally, we discuss the effect of an intrinsic spin relaxation on the dot which we model by adding a term $-\vec S/\tau$ to the right-hand side of the spin master equation, Eq.~\ref{eq:masterSpin}. Possible mechanisms for such a spin relaxation are the coupling to nuclear spins in the quantum dot~\cite{erlingsson_hyperfine-mediated_2002,merkulov_electron_2002,khaetskii_electron_2003} or spin-orbit interaction on the dot.~\cite{khaetskii_spin-flip_2001,golovach_phonon-induced_2004} In Fig.~\ref{fig:relaxation}, we show the current into the superconductor for different values of the relaxation time $\tau$. As the relaxation time is decreased, the current is reduced in agreement with our discussion above which showed that the spin accumulation on the dot is crucial to get a finite current. However, we also notice that the exchange field effects still remain visible when considering a finite relaxation. This shows that an experimental detection of these effects should be feasible.

\subsection{\label{ssec:asymmetry}Asymmetry effects}
\begin{figure}
	\includegraphics[width=\columnwidth]{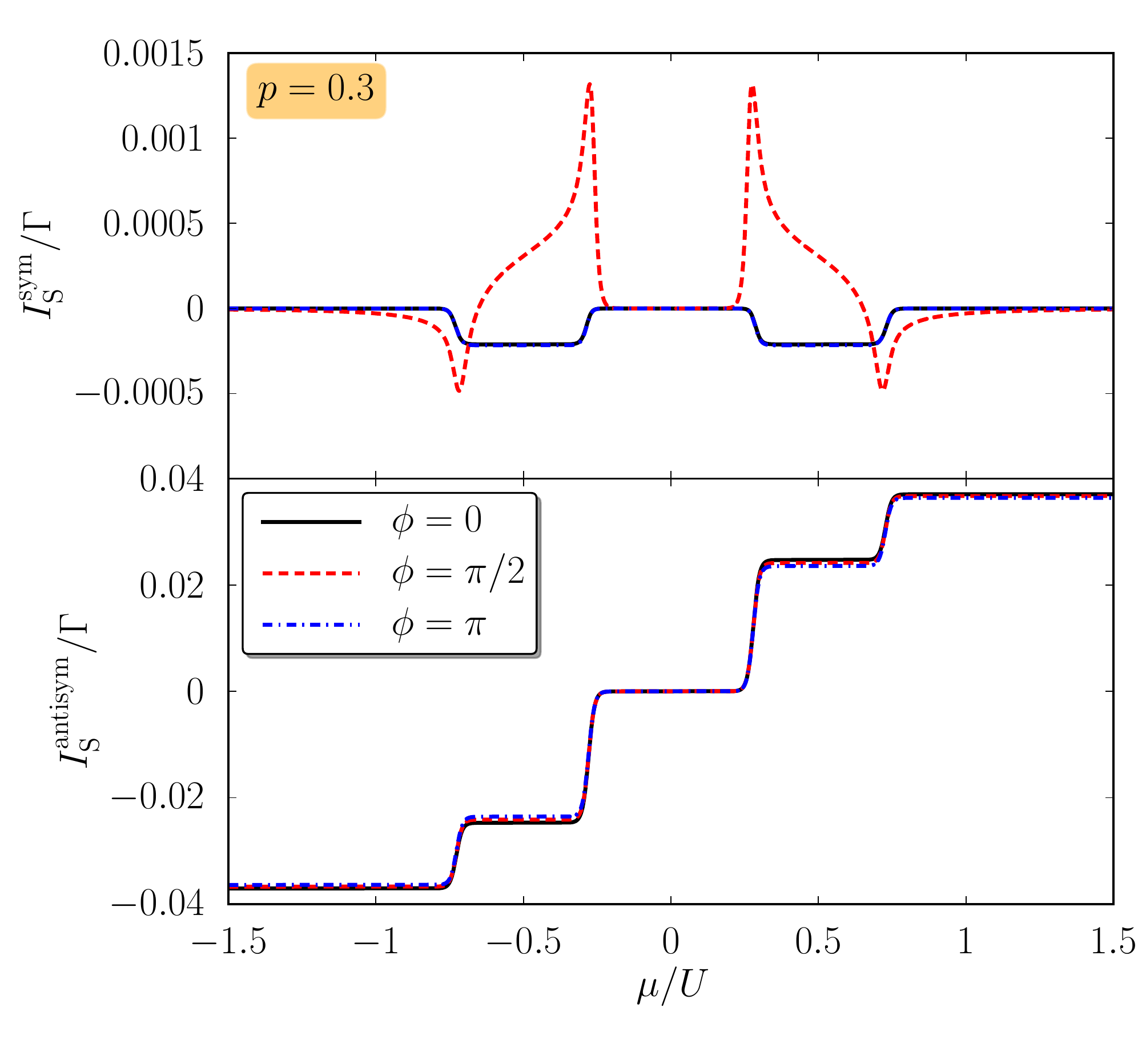}
	\includegraphics[width=\columnwidth]{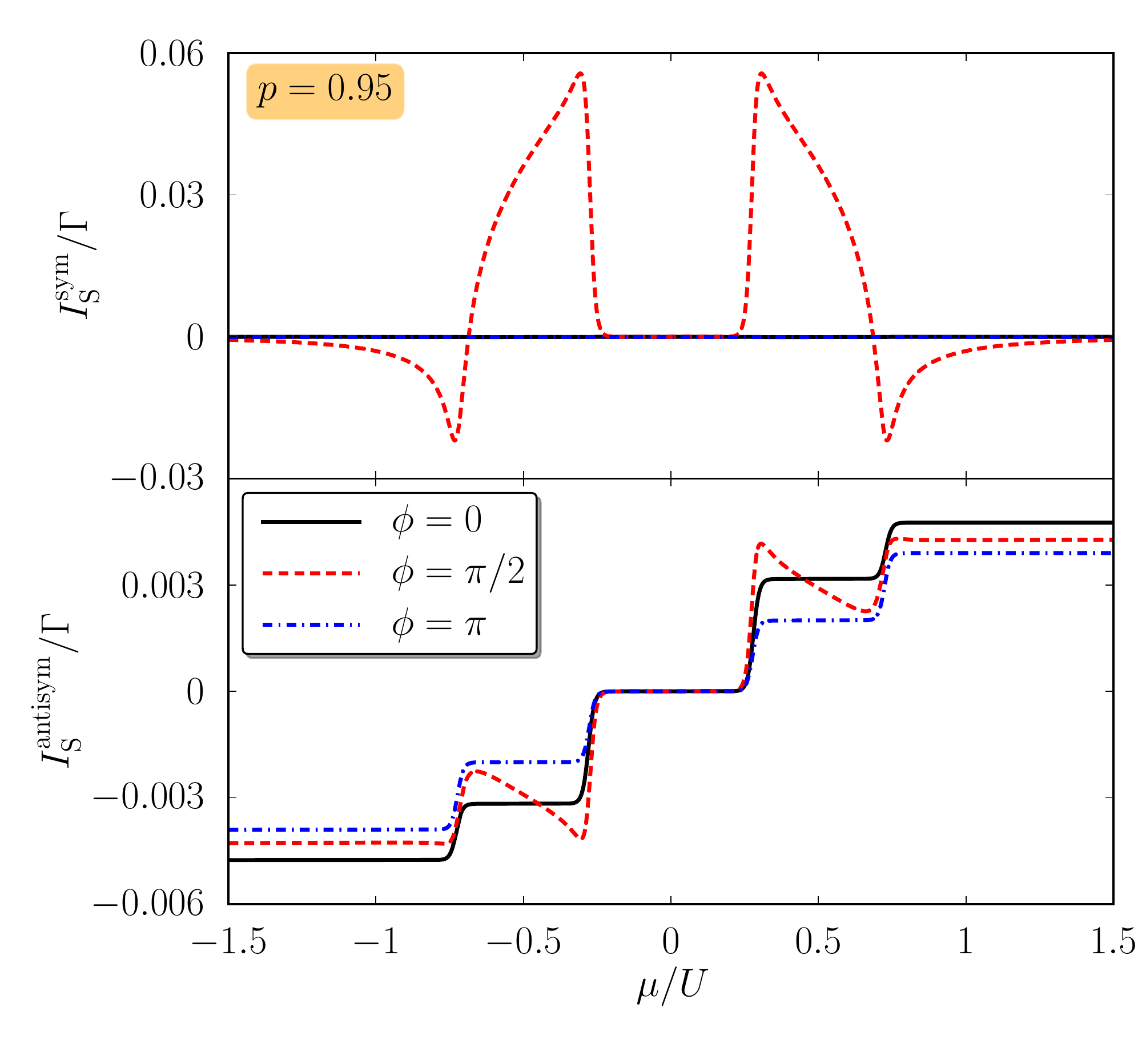}
	\caption{\label{fig:asymmetry}(Color online) Symmetrized and antisymmetrized current into the superconductor as a function of bias voltage for different magnetic configurations and small ($p=0.3$, upper panel) and large ($p=0.95$, lower panel) polarizations. The asymmetry is $a=0.05$, other parameters as in Fig.~\ref{fig:currentS}.}
\end{figure}
We now turn to the discussion of the situation where $\Gamma_\text{L}\neq\Gamma_\text{R}$. We parametrize the tunnel couplings as $\Gamma_\text{L}=(1+a)\Gamma/2$ and $\Gamma_\text{L}=(1-a)\Gamma/2$ such that the parameter $a$ with $-1\leq a\leq+1$ characterizes the degree of asymmetry. In this case, the antisymmetric component of the supercurrent with respect to the applied bias, $I_\text{S}^\text{antisym}(\mu)$, is in general non vanishing.

We find that for $a\neq0$ a finite supercurrent arises in the intermediate and large bias regime for all magnetic configurations. For parallel and antiparallel magnetizations magnetizations, respectively, the supercurrent is given by
\begin{align}
	I_\text{S}^\text{P}&=\Gamma\frac{2a\Gamma_\text{S}^2(1-p^2)(2\varepsilon_\text{A}-a\delta)}{\varepsilon_\text{A}\left[(3+a^2p^2)\Gamma_\text{S}^2-(1-p^2)\delta(a\varepsilon_\text{A}-(3-a^2)\delta)\right]}\\
	I_\text{S}^\text{AP}&=\Gamma\frac{a\Gamma_\text{S}^2(1-p^2)}{\varepsilon_\text{A}\left[(1+3p^2)a\delta+2(3+p^2)\varepsilon_\text{A}\right]}
\end{align}
in the intermediate bias regime, while it is given by
\begin{align}
	I_\text{S}^\text{P}&=\Gamma\frac{a(1-p^2)\Gamma_\text{S}^2}{(1-p^2)\delta^2+\Gamma_\text{S}^2},\\
	I_\text{S}^\text{AP}&=\Gamma\frac{a(1-p^2)\Gamma_\text{S}^2}{(1-a^2p^2)\delta^2+\Gamma_\text{S}^2}
\end{align}
in the large bias regime. In these formulas we assumed the current to flow from the left to the right. For a current in the opposite direction, one has to substitute $a\to-a$. Since the corresponding formulas for the noncollinear case are rather lengthy, we do not give them here.

From the above formulas we read off that the total supercurrent increases as the asymmetry is increased. Furthermore, we find that the current is decreased when the polarization is increased. This means that for experimentally realistic polarizations an asymmetry can give rise to a background current that dominates over the exchange-field signal. 
This problem can be overcome by looking at the symmetric and antisymmetric components of the current with respect to the voltage.

In Fig.~\ref{fig:asymmetry}, we show these two quantities as a function of applied bias for small and large polarizations. For small polarizations, we find that the antisymmetric contribution shows a steplike behavior that does not reveal any exchange field effects and is nearly insensitive to the magnetic configuration. In contrast, the symmetric contribution again reveals the characteristic peaks and dips that we encountered already in the symmetric system and that are a clear indication of the exchange field. For large polarizations, the system behaves rather similarly. The symmetric current contribution shows clear signs of the exchange field while the antisymmetric part is dominated by current steps. However, we now find that also the antisymmetric contribution is sensitive to the exchange field.

Hence, we have seen that an asymmetric coupling to the ferromagnets gives rise to a finite supercurrent for all magnetic configurations that could make the experimental detection of the exchange field effects difficult for small polarizations. To overcome this obstacle, we propose to measure the current symmetrized with respect to the bias voltage as this allows to recover the exchange field effects.

\section{\label{sec:conclusions}Conclusions}
We analyzed transport through a quantum-dot spin valve with an additional superconducting electrode. We find that in the case of noncollinear magnetization even for small polarizations, the symmetric component of the supercurrent  with respect to the applied bias voltage exhibits strong exchange-field effects.
In particular,  for a system that couples symmetrically to the ferromagnets which are at opposite bias, the supercurrent has only a symmetric component in bias voltage. In this case, a finite supercurrent can only flow for noncollinear magnetizations, as this configuration breaks the left-right symmetry for finite detuning $\delta$. Due to the presence of an exchange field acting on the dot spin in noncollinear geometries, the supercurrent exhibits a nontrivial bias dependence and even changes sign. Interestingly, these effects occur for any polarization of the ferromagnets $p>0$. Furthermore, they are robust towards a relaxation of the dot spin.
For systems with different couplings to the ferromagnets, the supercurrent becomes finite for any magnetic configuration. We find that for small polarizations, the contribution due to the asymmetry of the system  dominates over the one due to noncollinearity. We show, however, that by considering the supercurrent symmetrized with respect to the applied bias voltage, one can extract  the exchange field effects also in this case. We therefore propose to experimentally access the exchange field by measuring the bias dependence of the supercurrent.

\acknowledgments
We acknowledge financial support from DFG via SFB 491.

\appendix
\section{\label{sec:AppRules}Diagrammatic rules}
In this appendix, we give the diagrammatic rules (see also Ref.~\onlinecite{knig_zero-bias_1996,knig_resonant_1996,schoeller_transport_1997,knig_quantum_1999,braun_theory_2004,pala_nonequilibrium_2007,governale_real-time_2008,futterer_nonlocal_2009}) necessary to compute the kernels $W_{\chi_1\chi_1'}^{\chi_2\chi_2'}$ and $W_{\phantom{I_r}\chi_1\chi_1'}^{I_r\chi_2\chi_2'}$ that occur in Eq.~\eqref{eq:master} and Eq.~\eqref{eq:current} in the limit of an infinite superconducting gap, $\Delta\to\infty$.
\begin{enumerate}
	\item Draw all topological different diagrams with vertices on the propagators. Assign states $\chi$ and corresponding energies $E_\chi$ to the corners and all propagators. The vertices are contracted pairwise by tunneling lines that either conserve or flip the spin.
	\item Assign to all diagrams a resolvent $1/(\Delta E+i0^+)$ for each section on the contour between two adjacent vertices. Here $\Delta E$ is the energy difference between the left- and right-going propagators and tunneling lines.
	\item The tunneling lines involving the ferromagnetic electrode $r$ give rise to a factor of $\frac{\Gamma_r}{2\pi}f_r^\pm(\omega_i)$ if they do not flip the spin of the tunneling electron. If they flip it from up to down, they give rise to a factor of $\frac{p\Gamma}{2\pi}e^{i\phi_r}f_r^\pm(\omega_i)$. Flipping the spin in the opposite direction gives rise to the complex conjugate of the aforementioned factor. Here, the upper (lower) sign refers to lines running forward (backward) with respect to the Keldysh contour.
	\item Associate with each vertex that annihilates (creates) a dot electron with spin $\sigma$ a factor $\bra{\chi_2}c_\sigma\ket{\chi_1}$ ($\bra{\chi_2}c_\sigma^\dagger\ket{\chi_1}$). Here $\chi_1$ and $\chi_2$ are the states that enter and leave the vertex, respectively.
	\item Assign an overall prefactor $(-i)(-1)^{a+b}$ where $a$ is the number of vertices on the lower propagator and $b$ is the number of crossings of tunneling lines.
	\item Integrate over the energies of the tunneling lines $\omega_i$ and sum over all diagrams.
	\item To obtain the generalized current rates $W_{\phantom{I_r}\chi_1\chi_1'}^{I_r\chi_2\chi_2'}$  multiply each rate $W_{\chi_1\chi_1'}^{\chi_2\chi_2'}$ with a factor for all tunneling lines that are associated with lead $r$ that is the sum of the following numbers:
	\begin{enumerate}
		\item $+1$ if the line is going from the lower to the upper propagator,
		\item $-1$ if the line is going from the upper to the lower propagator,
		\item $0$ else.
	\end{enumerate}
\end{enumerate}

\section{\label{appendixb} Equation for the occupation probabilities}

In this Appendix we give the expressions for the quantities appearing in Eq.~\eqref{eq-diagonal} which can be obtained  as linear combinations of the kernels $W_{\chi_1\chi_1'}^{\chi_2\chi_2'}$ by reformulating the master equation~\eqref{eq:master} in terms of the occupation probabilities and average dot spin. We employ the abbreviations: $\Gamma_{r\pm}=\Gamma_r\left(1\pm\frac{\delta}{2\varepsilon_\text{A}}\right)$ and $f_{r\gamma\gamma'}^\pm=f_r^\pm(E_{\text{A},\gamma\gamma'})$, where $f_r^+(\omega)=1-f_r^-(\omega)$ denotes the Fermi function of lead $r$.

The matrix $\mathbf{A}_r$ depends on the position of the Andreev excitation energies and it reads
\begin{widetext}
\begin{equation}
\mathbf{A}_r	=
	\left(
	\begin{array}{ccc}
		-\Gamma_{r+}f_{r++}^--\Gamma_{r-}f_{r--}^+ & 0 & \frac{\Gamma_{r-}}{2}f_{r--}^-+\frac{\Gamma_{r+}}{2}f_{r++}^+ \\
		0 & -\Gamma_{r-}f_{r+-}^--\Gamma_{r+}f_{r-+}^+ & \frac{\Gamma_{r+}}{2}f_{r-+}^-+\frac{\Gamma_{r-}}{2}f_{r+-}^+ \\
		\Gamma_{r+}f_{r++}^-+\Gamma_{r-}f_{r--}^+ & \Gamma_{r-}f_{r+-}^-+\Gamma_{r+}f_{r-+}^+ & -\frac{\Gamma_{r-}}{2}f_{r--}^--\frac{\Gamma_{r+}}{2}f_{r-+}^--\frac{\Gamma_{r+}}{2}f_{r++}^+-\frac{\Gamma_{r-}}{2}f_{r+-}^+
	\end{array}
	\right).
\end{equation}
\end{widetext}

The vectors $\mathbf{b}_r$ describing the influence of the average spin on the dot on the diagonal probabilities are given by
\begin{equation}
\mathbf{b}_r=
	\left(
	\begin{array}{c}
		\Gamma_{r-}f_{r--}^--\Gamma_{r+}f_{r++}^+ \\
		\Gamma_{r+}f_{r-+}^--\Gamma_{r-}f_{r+-}^+ \\
		-\Gamma_{r-}f_{r--}^--\Gamma_{r+}f_{r-+}^-+\Gamma_{r+}f_{r++}^++\Gamma_{r-}f_{r+-}^+
	\end{array}
	\right).
\end{equation}


\begin{thebibliography}{10}%
\makeatletter
\providecommand \@ifxundefined [1]{%
 \ifx #1\undefined \expandafter \@firstoftwo
 \else \expandafter \@secondoftwo
\fi
}%
\providecommand \@ifnum [1]{%
 \ifnum #1\expandafter \@firstoftwo
 \else \expandafter \@secondoftwo
\fi
}%
\providecommand \enquote [1]{``#1''}%
\providecommand \bibnamefont  [1]{#1}%
\providecommand \bibfnamefont [1]{#1}%
\providecommand \citenamefont [1]{#1}%
\providecommand\href[0]{\@sanitize\@href}%
\providecommand\@href[1]{\endgroup\@@startlink{#1}\endgroup\@@href}%
\providecommand\@@href[1]{#1\@@endlink}%
\providecommand \@sanitize [0]{\begingroup\catcode`\&12\catcode`\#12\relax}%
\@ifxundefined \pdfoutput {\@firstoftwo}{%
 \@ifnum{\z@=\pdfoutput}{\@firstoftwo}{\@secondoftwo}%
}{%
 \providecommand\@@startlink[1]{\leavevmode}%
 \providecommand\@@endlink[0]{}%
}{%
 \providecommand\@@startlink[1]{%
  \leavevmode
  \pdfstartlink
   attr{/Border[0 0 1 ]/H/I/C[0 1 1]}%
   user{/Subtype/Link/A<</Type/Action/S/URI/URI(#1)>>}%
  \relax
 }%
 \providecommand\@@endlink[0]{\pdfendlink}%
}%
\providecommand \url  [0]{\begingroup\@sanitize \@url }%
\providecommand \@url [1]{\endgroup\@href {#1}{\urlprefix}}%
\providecommand \urlprefix [0]{URL }%
\providecommand \Eprint[0]{\href }%
\@ifxundefined \urlstyle {%
  \providecommand \doi [1]{doi:\discretionary{}{}{}#1}%
}{%
  \providecommand \doi [0]{doi:\discretionary{}{}{}\begingroup
  \urlstyle{rm}\Url }%
}%
\providecommand \doibase [0]{http://dx.doi.org/}%
\providecommand \Doi[1]{\href{\doibase#1}}%
\providecommand \bibAnnote [3]{%
  \BibitemShut{#1}%
  \begin{quotation}\noindent
    \textsc{Key:}\ #2\\\textsc{Annotation:}\ #3%
  \end{quotation}%
}%
\providecommand \bibAnnoteFile [2]{%
  \IfFileExists{#2}{\bibAnnote {#1} {#2} {\input{#2}}}{}%
}%
\providecommand \typeout [0]{\immediate \write \m@ne }%
\providecommand \selectlanguage [0]{\@gobble}%
\providecommand \bibinfo [0]{\@secondoftwo}%
\providecommand \bibfield [0]{\@secondoftwo}%
\providecommand \translation [1]{[#1]}%
\providecommand \BibitemOpen[0]{}%
\providecommand \bibitemStop [0]{}%
\providecommand \bibitemNoStop [0]{.\EOS\space}%
\providecommand \EOS [0]{\spacefactor3000\relax}%
\providecommand \BibitemShut [1]{\csname bibitem#1\endcsname}%
\bibitem{maekawa_spin_2002}%
  \BibitemOpen
  \bibfield{author}{%
  \bibinfo {author} {\bibfnamefont{S.}~\bibnamefont{Maekawa}}\ and\ \bibinfo
  {author} {\bibfnamefont{T.}~\bibnamefont{Shinjo}},\ }%
  \emph{\bibinfo {title} {Spin Dependent Transport in Magnetic
  Nanostructures}}\ (\bibinfo {publisher} {Taylor \& Francis},\ \bibinfo
  {address} {New York},\ \bibinfo {year} {2002})%
  \bibAnnoteFile{NoStop}{maekawa_spin_2002}%
\bibitem{maekawa_concepts_2006}%
  \BibitemOpen
  \bibfield{author}{%
  \bibinfo {author} {\bibfnamefont{S.}~\bibnamefont{Maekawa}},\ }%
  \emph{\bibinfo {title} {Concepts in Spin Electronics}}\ (\bibinfo {publisher}
  {Oxford University Press},\ \bibinfo {address} {New York},\ \bibinfo {year}
  {2006})%
  \bibAnnoteFile{NoStop}{maekawa_concepts_2006}%
\bibitem{deshmukh_using_2002}%
  \BibitemOpen
  \bibfield{author}{%
  \bibinfo {author} {\bibfnamefont{M.~M.}\ \bibnamefont{Deshmukh}}\ and\
  \bibinfo {author} {\bibfnamefont{D.~C.}\ \bibnamefont{Ralph}},\ }%
  \bibfield{journal}{%
  \Doi{10.1103/PhysRevLett.89.266803}{\bibinfo {journal} {Phys. Rev. Lett.}}\
  }%
  \textbf{\bibinfo {volume} {89}},\ \bibinfo {pages} {266803}\  (\bibinfo
  {year} {2002})%
  \bibAnnoteFile{NoStop}{deshmukh_using_2002}%
\bibitem{bernand-mantel_evidence_2006}%
  \BibitemOpen
  \bibfield{author}{%
  \bibinfo {author} {\bibfnamefont{A.}~\bibnamefont{{Bernand-Mantel}}},
  \bibinfo {author} {\bibfnamefont{P.}~\bibnamefont{Seneor}}, \bibinfo {author}
  {\bibfnamefont{N.}~\bibnamefont{Lidgi}}, \bibinfo {author}
  {\bibfnamefont{M.}~\bibnamefont{Munoz}}, \bibinfo {author}
  {\bibfnamefont{V.}~\bibnamefont{Cros}}, \bibinfo {author}
  {\bibfnamefont{S.}~\bibnamefont{Fusil}}, \bibinfo {author}
  {\bibfnamefont{K.}~\bibnamefont{Bouzehouane}}, \bibinfo {author}
  {\bibfnamefont{C.}~\bibnamefont{Deranlot}}, \bibinfo {author}
  {\bibfnamefont{A.}~\bibnamefont{Vaures}}, \bibinfo {author}
  {\bibfnamefont{F.}~\bibnamefont{Petroff}},\ and\ \bibinfo {author}
  {\bibfnamefont{A.}~\bibnamefont{Fert}},\ }%
  \bibfield{journal}{%
  \Doi{10.1063/1.2236293}{\bibinfo {journal} {Appl. Phys. Lett.}}\ }%
  \textbf{\bibinfo {volume} {89}},\ \bibinfo {pages} {062502}\  (\bibinfo
  {year} {2006})%
  \bibAnnoteFile{NoStop}{bernand-mantel_evidence_2006}%
\bibitem{mitani_current-induced_2008}%
  \BibitemOpen
  \bibfield{author}{%
  \bibinfo {author} {\bibfnamefont{S.}~\bibnamefont{Mitani}}, \bibinfo {author}
  {\bibfnamefont{Y.}~\bibnamefont{Nogi}}, \bibinfo {author}
  {\bibfnamefont{H.}~\bibnamefont{Wang}}, \bibinfo {author}
  {\bibfnamefont{K.}~\bibnamefont{Yakushiji}}, \bibinfo {author}
  {\bibfnamefont{F.}~\bibnamefont{Ernult}},\ and\ \bibinfo {author}
  {\bibfnamefont{K.}~\bibnamefont{Takanashi}},\ }%
  \bibfield{journal}{%
  \Doi{10.1063/1.2912036}{\bibinfo {journal} {Appl. Phys. Lett.}}\ }%
  \textbf{\bibinfo {volume} {92}},\ \bibinfo {pages} {152509}\  (\bibinfo
  {year} {2008})%
  \bibAnnoteFile{NoStop}{mitani_current-induced_2008}%
\bibitem{bernand-mantel_anisotropic_2009}%
  \BibitemOpen
  \bibfield{author}{%
  \bibinfo {author} {\bibfnamefont{A.}~\bibnamefont{{Bernand-Mantel}}},
  \bibinfo {author} {\bibfnamefont{P.}~\bibnamefont{Seneor}}, \bibinfo {author}
  {\bibfnamefont{K.}~\bibnamefont{Bouzehouane}}, \bibinfo {author}
  {\bibfnamefont{S.}~\bibnamefont{Fusil}}, \bibinfo {author}
  {\bibfnamefont{C.}~\bibnamefont{Deranlot}}, \bibinfo {author}
  {\bibfnamefont{F.}~\bibnamefont{Petroff}},\ and\ \bibinfo {author}
  {\bibfnamefont{A.}~\bibnamefont{Fert}},\ }%
  \bibfield{journal}{%
  \Doi{10.1038/nphys1423}{\bibinfo {journal} {Nat. Phys.}}\ }%
  \textbf{\bibinfo {volume} {5}},\ \bibinfo {pages} {920}\  (\bibinfo {year}
  {2009})%
  \bibAnnoteFile{NoStop}{bernand-mantel_anisotropic_2009}%
\bibitem{hamaya_spin_2007}%
  \BibitemOpen
  \bibfield{author}{%
  \bibinfo {author} {\bibfnamefont{K.}~\bibnamefont{Hamaya}}, \bibinfo {author}
  {\bibfnamefont{S.}~\bibnamefont{Masubuchi}}, \bibinfo {author}
  {\bibfnamefont{M.}~\bibnamefont{Kawamura}}, \bibinfo {author}
  {\bibfnamefont{T.}~\bibnamefont{Machida}}, \bibinfo {author}
  {\bibfnamefont{M.}~\bibnamefont{Jung}}, \bibinfo {author}
  {\bibfnamefont{K.}~\bibnamefont{Shibata}}, \bibinfo {author}
  {\bibfnamefont{K.}~\bibnamefont{Hirakawa}}, \bibinfo {author}
  {\bibfnamefont{T.}~\bibnamefont{Taniyama}}, \bibinfo {author}
  {\bibfnamefont{S.}~\bibnamefont{Ishida}},\ and\ \bibinfo {author}
  {\bibfnamefont{Y.}~\bibnamefont{Arakawa}},\ }%
  \bibfield{journal}{%
  \Doi{10.1063/1.2435957}{\bibinfo {journal} {Appl. Phys. Lett.}}\ }%
  \textbf{\bibinfo {volume} {90}},\ \bibinfo {pages} {053108}\  (\bibinfo
  {year} {2007})%
  \bibAnnoteFile{NoStop}{hamaya_spin_2007}%
\bibitem{hamaya_electric-field_2007}%
  \BibitemOpen
  \bibfield{author}{%
  \bibinfo {author} {\bibfnamefont{K.}~\bibnamefont{Hamaya}}, \bibinfo {author}
  {\bibfnamefont{M.}~\bibnamefont{Kitabatake}}, \bibinfo {author}
  {\bibfnamefont{K.}~\bibnamefont{Shibata}}, \bibinfo {author}
  {\bibfnamefont{M.}~\bibnamefont{Jung}}, \bibinfo {author}
  {\bibfnamefont{M.}~\bibnamefont{Kawamura}}, \bibinfo {author}
  {\bibfnamefont{K.}~\bibnamefont{Hirakawa}}, \bibinfo {author}
  {\bibfnamefont{T.}~\bibnamefont{Machida}}, \bibinfo {author}
  {\bibfnamefont{T.}~\bibnamefont{Taniyama}}, \bibinfo {author}
  {\bibfnamefont{S.}~\bibnamefont{Ishida}},\ and\ \bibinfo {author}
  {\bibfnamefont{Y.}~\bibnamefont{Arakawa}},\ }%
  \bibfield{journal}{%
  \Doi{10.1063/1.2759264}{\bibinfo {journal} {Appl. Phys. Lett.}}\ }%
  \textbf{\bibinfo {volume} {91}},\ \bibinfo {pages} {022107}\  (\bibinfo
  {year} {2007})%
  \bibAnnoteFile{NoStop}{hamaya_electric-field_2007}%
\bibitem{hamaya_kondo_2007}%
  \BibitemOpen
  \bibfield{author}{%
  \bibinfo {author} {\bibfnamefont{K.}~\bibnamefont{Hamaya}}, \bibinfo {author}
  {\bibfnamefont{M.}~\bibnamefont{Kitabatake}}, \bibinfo {author}
  {\bibfnamefont{K.}~\bibnamefont{Shibata}}, \bibinfo {author}
  {\bibfnamefont{M.}~\bibnamefont{Jung}}, \bibinfo {author}
  {\bibfnamefont{M.}~\bibnamefont{Kawamura}}, \bibinfo {author}
  {\bibfnamefont{K.}~\bibnamefont{Hirakawa}}, \bibinfo {author}
  {\bibfnamefont{T.}~\bibnamefont{Machida}}, \bibinfo {author}
  {\bibfnamefont{T.}~\bibnamefont{Taniyama}}, \bibinfo {author}
  {\bibfnamefont{S.}~\bibnamefont{Ishida}},\ and\ \bibinfo {author}
  {\bibfnamefont{Y.}~\bibnamefont{Arakawa}},\ }%
  \bibfield{journal}{%
  \Doi{10.1063/1.2820445}{\bibinfo {journal} {Appl. Phys. Lett.}}\ }%
  \textbf{\bibinfo {volume} {91}},\ \bibinfo {pages} {232105}\  (\bibinfo
  {year} {2007})%
  \bibAnnoteFile{NoStop}{hamaya_kondo_2007}%
\bibitem{hamaya_oscillatory_2008}%
  \BibitemOpen
  \bibfield{author}{%
  \bibinfo {author} {\bibfnamefont{K.}~\bibnamefont{Hamaya}}, \bibinfo {author}
  {\bibfnamefont{M.}~\bibnamefont{Kitabatake}}, \bibinfo {author}
  {\bibfnamefont{K.}~\bibnamefont{Shibata}}, \bibinfo {author}
  {\bibfnamefont{M.}~\bibnamefont{Jung}}, \bibinfo {author}
  {\bibfnamefont{M.}~\bibnamefont{Kawamura}}, \bibinfo {author}
  {\bibfnamefont{S.}~\bibnamefont{Ishida}}, \bibinfo {author}
  {\bibfnamefont{T.}~\bibnamefont{Taniyama}}, \bibinfo {author}
  {\bibfnamefont{K.}~\bibnamefont{Hirakawa}}, \bibinfo {author}
  {\bibfnamefont{Y.}~\bibnamefont{Arakawa}},\ and\ \bibinfo {author}
  {\bibfnamefont{T.}~\bibnamefont{Machida}},\ }%
  \bibfield{journal}{%
  \Doi{10.1103/PhysRevB.77.081302}{\bibinfo {journal} {Phys. Rev. B.}}\ }%
  \textbf{\bibinfo {volume} {77}},\ \bibinfo {pages} {081302}\  (\bibinfo
  {year} {2008})%
  \bibAnnoteFile{NoStop}{hamaya_oscillatory_2008}%
\bibitem{hamaya_tunneling_2008}%
  \BibitemOpen
  \bibfield{author}{%
  \bibinfo {author} {\bibfnamefont{K.}~\bibnamefont{Hamaya}}, \bibinfo {author}
  {\bibfnamefont{M.}~\bibnamefont{Kitabatake}}, \bibinfo {author}
  {\bibfnamefont{K.}~\bibnamefont{Shibata}}, \bibinfo {author}
  {\bibfnamefont{M.}~\bibnamefont{Jung}}, \bibinfo {author}
  {\bibfnamefont{M.}~\bibnamefont{Kawamura}}, \bibinfo {author}
  {\bibfnamefont{S.}~\bibnamefont{Ishida}}, \bibinfo {author}
  {\bibfnamefont{T.}~\bibnamefont{Taniyama}}, \bibinfo {author}
  {\bibfnamefont{K.}~\bibnamefont{Hirakawa}}, \bibinfo {author}
  {\bibfnamefont{Y.}~\bibnamefont{Arakawa}},\ and\ \bibinfo {author}
  {\bibfnamefont{T.}~\bibnamefont{Machida}},\ }%
  \bibfield{journal}{%
  \Doi{10.1063/1.3042098}{\bibinfo {journal} {Appl. Phys. Lett.}}\ }%
  \textbf{\bibinfo {volume} {93}},\ \bibinfo {pages} {222107}\  (\bibinfo
  {year} {2008})%
  \bibAnnoteFile{NoStop}{hamaya_tunneling_2008}%
\bibitem{hofstetter_ferromagnetic_2010}%
  \BibitemOpen
  \bibfield{author}{%
  \bibinfo {author} {\bibfnamefont{L.}~\bibnamefont{Hofstetter}}, \bibinfo
  {author} {\bibfnamefont{A.}~\bibnamefont{Geresdi}}, \bibinfo {author}
  {\bibfnamefont{M.}~\bibnamefont{Aagesen}}, \bibinfo {author}
  {\bibfnamefont{J.}~\bibnamefont{Nygard}}, \bibinfo {author}
  {\bibfnamefont{C.}~\bibnamefont{Sch\"onenberger}},\ and\ \bibinfo {author}
  {\bibfnamefont{S.}~\bibnamefont{Csonka}},\ }%
  \bibfield{journal}{%
  \Doi{10.1103/PhysRevLett.104.246804}{\bibinfo {journal} {Phys. Rev. Lett.}}\
  }%
  \textbf{\bibinfo {volume} {104}},\ \bibinfo {pages} {246804}\  (\bibinfo
  {year} {2010})%
  \bibAnnoteFile{NoStop}{hofstetter_ferromagnetic_2010}%
\bibitem{jensen_single-wall_2003}%
  \BibitemOpen
  \bibfield{author}{%
  \bibinfo {author} {\bibfnamefont{A.}~\bibnamefont{Jensen}}, \bibinfo {author}
  {\bibfnamefont{J.}~\bibnamefont{Nygård}},\ and\ \bibinfo {author}
  {\bibfnamefont{J.}~\bibnamefont{Borggreen}},\ }%
  in\ \emph{\bibinfo {booktitle} {Proceedings of the International Symposium on
  Mesoscopic Superconductivity and Spintronics}},\ \bibinfo {editor} {edited
  by\ \bibinfo {editor} {\bibfnamefont{H.}~\bibnamefont{Takayanagi}}\ and\
  \bibinfo {editor} {\bibfnamefont{J.}~\bibnamefont{Nitta}}}\ (\bibinfo
  {publisher} {World Scientific},\ \bibinfo {address} {Singapore},\ \bibinfo
  {year} {2003})\ pp.\ \bibinfo {pages} {33--37}%
  \bibAnnoteFile{NoStop}{jensen_single-wall_2003}%
\bibitem{sahoo_electric_2005}%
  \BibitemOpen
  \bibfield{author}{%
  \bibinfo {author} {\bibfnamefont{S.}~\bibnamefont{Sahoo}}, \bibinfo {author}
  {\bibfnamefont{T.}~\bibnamefont{Kontos}}, \bibinfo {author}
  {\bibfnamefont{J.}~\bibnamefont{Furer}}, \bibinfo {author}
  {\bibfnamefont{C.}~\bibnamefont{Hoffmann}}, \bibinfo {author}
  {\bibfnamefont{M.}~\bibnamefont{Graber}}, \bibinfo {author}
  {\bibfnamefont{A.}~\bibnamefont{Cottet}},\ and\ \bibinfo {author}
  {\bibfnamefont{C.}~\bibnamefont{Sch\"onenberger}},\ }%
  \bibfield{journal}{%
  \Doi{10.1038/nphys149}{\bibinfo {journal} {Nat. Phys.}}\ }%
  \textbf{\bibinfo {volume} {1}},\ \bibinfo {pages} {99}\  (\bibinfo {year}
  {2005})%
  \bibAnnoteFile{NoStop}{sahoo_electric_2005}%
\bibitem{jensen_hybrid_2004}%
  \BibitemOpen
  \bibfield{author}{%
  \bibinfo {author} {\bibfnamefont{A.}~\bibnamefont{Jensen}}, \bibinfo {author}
  {\bibfnamefont{J.~R.}\ \bibnamefont{Hauptmann}}, \bibinfo {author}
  {\bibfnamefont{J.}~\bibnamefont{Nygård}}, \bibinfo {author}
  {\bibfnamefont{J.}~\bibnamefont{Sadowski}},\ and\ \bibinfo {author}
  {\bibfnamefont{P.~E.}\ \bibnamefont{Lindelof}},\ }%
  \bibfield{journal}{%
  \Doi{10.1021/nl0350027}{\bibinfo {journal} {Nano Lett.}}\ }%
  \textbf{\bibinfo {volume} {4}},\ \bibinfo {pages} {349}\  (\bibinfo {year}
  {2004})%
  \bibAnnoteFile{NoStop}{jensen_hybrid_2004}%
\bibitem{hauptmann_electric-field-controlled_2008}%
  \BibitemOpen
  \bibfield{author}{%
  \bibinfo {author} {\bibfnamefont{J.~R.}\ \bibnamefont{Hauptmann}}, \bibinfo
  {author} {\bibfnamefont{J.}~\bibnamefont{Paaske}},\ and\ \bibinfo {author}
  {\bibfnamefont{P.~E.}\ \bibnamefont{Lindelof}},\ }%
  \bibfield{journal}{%
  \Doi{10.1038/nphys931}{\bibinfo {journal} {Nat. Phys.}}\ }%
  \textbf{\bibinfo {volume} {4}},\ \bibinfo {pages} {373}\  (\bibinfo {year}
  {2008})%
  \bibAnnoteFile{NoStop}{hauptmann_electric-field-controlled_2008}%
\bibitem{merchant_current_2009}%
  \BibitemOpen
  \bibfield{author}{%
  \bibinfo {author} {\bibfnamefont{C.~A.}\ \bibnamefont{Merchant}}\ and\
  \bibinfo {author} {\bibfnamefont{N.}~\bibnamefont{Markovic}},\ }%
  in\ \Doi{10.1063/1.3072020}{\emph{\bibinfo {booktitle} {Proceedings of the
  53rd annual conference on magnetism and magnetic materials}}},\ Vol.\
  \bibinfo {volume} {105}\ (\bibinfo {publisher} {{AIP}},\ \bibinfo {address}
  {Austin, Texas {(USA)}},\ \bibinfo {year} {2009})\ \ p.\ \bibinfo {pages}
  {07C711}%
  \bibAnnoteFile{NoStop}{merchant_current_2009}%
\bibitem{pasupathy_kondo_2004}%
  \BibitemOpen
  \bibfield{author}{%
  \bibinfo {author} {\bibfnamefont{A.~N.}\ \bibnamefont{Pasupathy}}, \bibinfo
  {author} {\bibfnamefont{R.~C.}\ \bibnamefont{Bialczak}}, \bibinfo {author}
  {\bibfnamefont{J.}~\bibnamefont{Martinek}}, \bibinfo {author}
  {\bibfnamefont{J.~E.}\ \bibnamefont{Grose}}, \bibinfo {author}
  {\bibfnamefont{L.~A.~K.}\ \bibnamefont{Donev}}, \bibinfo {author}
  {\bibfnamefont{P.~L.}\ \bibnamefont{{McEuen}}},\ and\ \bibinfo {author}
  {\bibfnamefont{D.~C.}\ \bibnamefont{Ralph}},\ }%
  \bibfield{journal}{%
  \Doi{10.1126/science.1102068}{\bibinfo {journal} {Science}}\ }%
  \textbf{\bibinfo {volume} {306}},\ \bibinfo {pages} {86}\  (\bibinfo {year}
  {2004})%
  \bibAnnoteFile{NoStop}{pasupathy_kondo_2004}%
\bibitem{souza_quantum_2007}%
  \BibitemOpen
  \bibfield{author}{%
  \bibinfo {author} {\bibfnamefont{F.~M.}\ \bibnamefont{Souza}}, \bibinfo
  {author} {\bibfnamefont{J.~C.}\ \bibnamefont{Egues}},\ and\ \bibinfo {author}
  {\bibfnamefont{A.~P.}\ \bibnamefont{Jauho}},\ }%
  \bibfield{journal}{%
  \Doi{10.1103/PhysRevB.75.165303}{\bibinfo {journal} {Phys. Rev. B.}}\ }%
  \textbf{\bibinfo {volume} {75}},\ \bibinfo {pages} {165303}\  (\bibinfo
  {year} {2007})%
  \bibAnnoteFile{NoStop}{souza_quantum_2007}%
\bibitem{weymann_spin_2008}%
  \BibitemOpen
  \bibfield{author}{%
  \bibinfo {author} {\bibfnamefont{I.}~\bibnamefont{Weymann}}\ and\ \bibinfo
  {author} {\bibfnamefont{J.}~\bibnamefont{Barnas}},\ }%
  \bibfield{journal}{%
  \Doi{10.1063/1.2894224}{\bibinfo {journal} {Appl. Phys. Lett.}}\ }%
  \textbf{\bibinfo {volume} {92}},\ \bibinfo {pages} {103127}\  (\bibinfo
  {year} {2008})%
  \bibAnnoteFile{NoStop}{weymann_spin_2008}%
\bibitem{knig_interaction-driven_2003}%
  \BibitemOpen
  \bibfield{author}{%
  \bibinfo {author} {\bibfnamefont{J.}~\bibnamefont{K\"onig}}\ and\ \bibinfo
  {author} {\bibfnamefont{J.}~\bibnamefont{Martinek}},\ }%
  \bibfield{journal}{%
  \Doi{10.1103/PhysRevLett.90.166602}{\bibinfo {journal} {Phys. Rev. Lett.}}\
  }%
  \textbf{\bibinfo {volume} {90}},\ \bibinfo {pages} {166602}\  (\bibinfo
  {year} {2003})%
  \bibAnnoteFile{NoStop}{knig_interaction-driven_2003}%
\bibitem{braun_theory_2004}%
  \BibitemOpen
  \bibfield{author}{%
  \bibinfo {author} {\bibfnamefont{M.}~\bibnamefont{Braun}}, \bibinfo {author}
  {\bibfnamefont{J.}~\bibnamefont{K\"onig}},\ and\ \bibinfo {author}
  {\bibfnamefont{J.}~\bibnamefont{Martinek}},\ }%
  \bibfield{journal}{%
  \Doi{10.1103/PhysRevB.70.195345}{\bibinfo {journal} {Phys. Rev. B.}}\ }%
  \textbf{\bibinfo {volume} {70}},\ \bibinfo {pages} {195345}\  (\bibinfo
  {year} {2004})%
  \bibAnnoteFile{NoStop}{braun_theory_2004}%
\bibitem{fransson_angular_2005-1}%
  \BibitemOpen
  \bibfield{author}{%
  \bibinfo {author} {\bibfnamefont{J.}~\bibnamefont{Fransson}},\ }%
  \bibfield{journal}{%
  \Doi{10.1103/PhysRevB.72.045415}{\bibinfo {journal} {Phys. Rev. B.}}\ }%
  \textbf{\bibinfo {volume} {72}},\ \bibinfo {pages} {045415}\  (\bibinfo
  {year} {2005})%
  \bibAnnoteFile{NoStop}{fransson_angular_2005-1}%
\bibitem{pedersen_noncollinear_2005}%
  \BibitemOpen
  \bibfield{author}{%
  \bibinfo {author} {\bibfnamefont{J.~N.}\ \bibnamefont{Pedersen}}, \bibinfo
  {author} {\bibfnamefont{J.~Q.}\ \bibnamefont{Thomassen}},\ and\ \bibinfo
  {author} {\bibfnamefont{K.}~\bibnamefont{Flensberg}},\ }%
  \bibfield{journal}{%
  \Doi{10.1103/PhysRevB.72.045341}{\bibinfo {journal} {Phys. Rev. B.}}\ }%
  \textbf{\bibinfo {volume} {72}},\ \bibinfo {pages} {045341}\  (\bibinfo
  {year} {2005})%
  \bibAnnoteFile{NoStop}{pedersen_noncollinear_2005}%
\bibitem{fransson_angular_2005}%
  \BibitemOpen
  \bibfield{author}{%
  \bibinfo {author} {\bibfnamefont{J.}~\bibnamefont{Fransson}},\ }%
  \bibfield{journal}{%
  \Doi{10.1209/epl/i2005-10043-1}{\bibinfo {journal} {Europhys. Lett.}}\ }%
  \textbf{\bibinfo {volume} {70}},\ \bibinfo {pages} {796}\  (\bibinfo {year}
  {2005})%
  \bibAnnoteFile{NoStop}{fransson_angular_2005}%
\bibitem{weymann_cotunneling_2005}%
  \BibitemOpen
  \bibfield{author}{%
  \bibinfo {author} {\bibfnamefont{I.}~\bibnamefont{Weymann}}\ and\ \bibinfo
  {author} {\bibfnamefont{J.}~\bibnamefont{Barnas}},\ }%
  \bibfield{journal}{%
  \Doi{10.1140/epjb/e2005-00227-y}{\bibinfo {journal} {Eur. Phys. Journ. B}}\
  }%
  \textbf{\bibinfo {volume} {46}},\ \bibinfo {pages} {289}\  (\bibinfo {year}
  {2005})%
  \bibAnnoteFile{NoStop}{weymann_cotunneling_2005}%
\bibitem{weymann_cotunneling_2007}%
  \BibitemOpen
  \bibfield{author}{%
  \bibinfo {author} {\bibfnamefont{I.}~\bibnamefont{Weymann}}\ and\ \bibinfo
  {author} {\bibfnamefont{J.}~\bibnamefont{Barnas}},\ }%
  \bibfield{journal}{%
  \Doi{10.1103/PhysRevB.75.155308}{\bibinfo {journal} {Phys. Rev. B.}}\ }%
  \textbf{\bibinfo {volume} {75}},\ \bibinfo {pages} {155308}\  (\bibinfo
  {year} {2007})%
  \bibAnnoteFile{NoStop}{weymann_cotunneling_2007}%
\bibitem{koller_spin_2007}%
  \BibitemOpen
  \bibfield{author}{%
  \bibinfo {author} {\bibfnamefont{S.}~\bibnamefont{Koller}}, \bibinfo {author}
  {\bibfnamefont{L.}~\bibnamefont{Mayrhofer}},\ and\ \bibinfo {author}
  {\bibfnamefont{M.}~\bibnamefont{Grifoni}},\ }%
  \bibfield{journal}{%
  \Doi{10.1088/1367-2630/9/9/348}{\bibinfo {journal} {New J. Phys.}}\ }%
  \textbf{\bibinfo {volume} {9}},\ \bibinfo {pages} {348}\  (\bibinfo {year}
  {2007})%
  \bibAnnoteFile{NoStop}{koller_spin_2007}%
\bibitem{cottet_magnetoresistance_2006}%
  \BibitemOpen
  \bibfield{author}{%
  \bibinfo {author} {\bibfnamefont{A.}~\bibnamefont{Cottet}}\ and\ \bibinfo
  {author} {\bibfnamefont{M.~S.}~\bibnamefont{Choi}},\ }%
  \bibfield{journal}{%
  \Doi{10.1103/PhysRevB.74.235316}{\bibinfo {journal} {Phys. Rev. B.}}\ }%
  \textbf{\bibinfo {volume} {74}},\ \bibinfo {pages} {235316}\  (\bibinfo
  {year} {2006})%
  \bibAnnoteFile{NoStop}{cottet_magnetoresistance_2006}%
\bibitem{trocha_quantum_2007}%
  \BibitemOpen
  \bibfield{author}{%
  \bibinfo {author} {\bibfnamefont{P.}~\bibnamefont{Trocha}}\ and\ \bibinfo
  {author} {\bibfnamefont{J.}~\bibnamefont{Barnas}},\ }%
  \bibfield{journal}{%
  \Doi{10.1103/PhysRevB.76.165432}{\bibinfo {journal} {Phys. Rev. B.}}\ }%
  \textbf{\bibinfo {volume} {76}},\ \bibinfo {pages} {165432}\  (\bibinfo
  {year} {2007})%
  \bibAnnoteFile{NoStop}{trocha_quantum_2007}%
\bibitem{hornberger_transport_2008}%
  \BibitemOpen
  \bibfield{author}{%
  \bibinfo {author} {\bibfnamefont{R.}~\bibnamefont{Hornberger}}, \bibinfo
  {author} {\bibfnamefont{S.}~\bibnamefont{Koller}}, \bibinfo {author}
  {\bibfnamefont{G.}~\bibnamefont{Begemann}}, \bibinfo {author}
  {\bibfnamefont{A.}~\bibnamefont{Donarini}},\ and\ \bibinfo {author}
  {\bibfnamefont{M.}~\bibnamefont{Grifoni}},\ }%
  \bibfield{journal}{%
  \Doi{10.1103/PhysRevB.77.245313}{\bibinfo {journal} {Phys. Rev. B.}}\ }%
  \textbf{\bibinfo {volume} {77}},\ \bibinfo {pages} {245313}\  (\bibinfo
  {year} {2008})%
  \bibAnnoteFile{NoStop}{hornberger_transport_2008}%
\bibitem{trocha_negative_2009}%
  \BibitemOpen
  \bibfield{author}{%
  \bibinfo {author} {\bibfnamefont{P.}~\bibnamefont{Trocha}}, \bibinfo {author}
  {\bibfnamefont{I.}~\bibnamefont{Weymann}},\ and\ \bibinfo {author}
  {\bibfnamefont{J.}~\bibnamefont{Barnas}},\ }%
  \bibfield{journal}{%
  \Doi{10.1103/PhysRevB.80.165333}{\bibinfo {journal} {Phys. Rev. B.}}\ }%
  \textbf{\bibinfo {volume} {80}},\ \bibinfo {pages} {165333}\  (\bibinfo
  {year} {2009})%
  \bibAnnoteFile{NoStop}{trocha_negative_2009}%
\bibitem{buka_current_2000}%
  \BibitemOpen
  \bibfield{author}{%
  \bibinfo {author} {\bibfnamefont{B.~R.}\ \bibnamefont{Bułka}},\ }%
  \bibfield{journal}{%
  \Doi{10.1103/PhysRevB.62.1186}{\bibinfo {journal} {Phys. Rev. B.}}\ }%
  \textbf{\bibinfo {volume} {62}},\ \bibinfo {pages} {1186}\  (\bibinfo {year}
  {2000})%
  \bibAnnoteFile{NoStop}{buka_current_2000}%
\bibitem{weymann_theory_2007}%
  \BibitemOpen
  \bibfield{author}{%
  \bibinfo {author} {\bibfnamefont{I.}~\bibnamefont{Weymann}}, \bibinfo
  {author} {\bibfnamefont{J.}~\bibnamefont{Barnas}},\ and\ \bibinfo {author}
  {\bibfnamefont{S.}~\bibnamefont{Krompiewski}},\ }%
  \bibfield{journal}{%
  \Doi{10.1103/PhysRevB.76.155408}{\bibinfo {journal} {Phys. Rev. B.}}\ }%
  \textbf{\bibinfo {volume} {76}},\ \bibinfo {pages} {155408}\  (\bibinfo
  {year} {2007})%
  \bibAnnoteFile{NoStop}{weymann_theory_2007}%
\bibitem{weymann_shot_2008}%
  \BibitemOpen
  \bibfield{author}{%
  \bibinfo {author} {\bibfnamefont{I.}~\bibnamefont{Weymann}}\ and\ \bibinfo
  {author} {\bibfnamefont{J.}~\bibnamefont{Barnas}},\ }%
  \bibfield{journal}{%
  \Doi{10.1103/PhysRevB.77.075305}{\bibinfo {journal} {Phys. Rev. B.}}\ }%
  \textbf{\bibinfo {volume} {77}},\ \bibinfo {pages} {075305}\  (\bibinfo
  {year} {2008})%
  \bibAnnoteFile{NoStop}{weymann_shot_2008}%
\bibitem{souza_spin-polarized_2008}%
  \BibitemOpen
  \bibfield{author}{%
  \bibinfo {author} {\bibfnamefont{F.~M.}\ \bibnamefont{Souza}}, \bibinfo
  {author} {\bibfnamefont{A.~P.}\ \bibnamefont{Jauho}},\ and\ \bibinfo {author}
  {\bibfnamefont{J.~C.}\ \bibnamefont{Egues}},\ }%
  \bibfield{journal}{%
  \Doi{10.1103/PhysRevB.78.155303}{\bibinfo {journal} {Phys. Rev. B.}}\ }%
  \textbf{\bibinfo {volume} {78}},\ \bibinfo {pages} {155303}\  (\bibinfo
  {year} {2008})%
  \bibAnnoteFile{NoStop}{souza_spin-polarized_2008}%
\bibitem{lindebaum_spin-induced_2009}%
  \BibitemOpen
  \bibfield{author}{%
  \bibinfo {author} {\bibfnamefont{S.}~\bibnamefont{Lindebaum}}, \bibinfo
  {author} {\bibfnamefont{D.}~\bibnamefont{Urban}},\ and\ \bibinfo {author}
  {\bibfnamefont{J.}~\bibnamefont{K\"onig}},\ }%
  \bibfield{journal}{%
  \Doi{10.1103/PhysRevB.79.245303}{\bibinfo {journal} {Phys. Rev. B.}}\ }%
  \textbf{\bibinfo {volume} {79}},\ \bibinfo {pages} {245303}\  (\bibinfo
  {year} {2009})%
  \bibAnnoteFile{NoStop}{lindebaum_spin-induced_2009}%
\bibitem{braun_frequency-dependent_2006}%
  \BibitemOpen
  \bibfield{author}{%
  \bibinfo {author} {\bibfnamefont{M.}~\bibnamefont{Braun}}, \bibinfo {author}
  {\bibfnamefont{J.}~\bibnamefont{K\"onig}},\ and\ \bibinfo {author}
  {\bibfnamefont{J.}~\bibnamefont{Martinek}},\ }%
  \bibfield{journal}{%
  \Doi{10.1103/PhysRevB.74.075328}{\bibinfo {journal} {Phys. Rev. B.}}\ }%
  \textbf{\bibinfo {volume} {74}},\ \bibinfo {pages} {075328}\  (\bibinfo
  {year} {2006})%
  \bibAnnoteFile{NoStop}{braun_frequency-dependent_2006}%
\bibitem{martinek_kondo_2003}%
  \BibitemOpen
  \bibfield{author}{%
  \bibinfo {author} {\bibfnamefont{J.}~\bibnamefont{Martinek}}, \bibinfo
  {author} {\bibfnamefont{Y.}~\bibnamefont{Utsumi}}, \bibinfo {author}
  {\bibfnamefont{H.}~\bibnamefont{Imamura}}, \bibinfo {author}
  {\bibfnamefont{J.}~\bibnamefont{Barnas}}, \bibinfo {author}
  {\bibfnamefont{S.}~\bibnamefont{Maekawa}}, \bibinfo {author}
  {\bibfnamefont{J.}~\bibnamefont{K\"onig}},\ and\ \bibinfo {author}
  {\bibfnamefont{G.}~\bibnamefont{Sch\"on}},\ }%
  \bibfield{journal}{%
  \Doi{10.1103/PhysRevLett.91.127203}{\bibinfo {journal} {Phys. Rev. Lett.}}\
  }%
  \textbf{\bibinfo {volume} {91}},\ \bibinfo {pages} {127203}\  (\bibinfo
  {year} {2003})%
  \bibAnnoteFile{NoStop}{martinek_kondo_2003}%
\bibitem{martinek_kondo_2003-1}%
  \BibitemOpen
  \bibfield{author}{%
  \bibinfo {author} {\bibfnamefont{J.}~\bibnamefont{Martinek}}, \bibinfo
  {author} {\bibfnamefont{M.}~\bibnamefont{Sindel}}, \bibinfo {author}
  {\bibfnamefont{L.}~\bibnamefont{Borda}}, \bibinfo {author}
  {\bibfnamefont{J.}~\bibnamefont{Barnas}}, \bibinfo {author}
  {\bibfnamefont{J.}~\bibnamefont{K\"onig}}, \bibinfo {author}
  {\bibfnamefont{G.}~\bibnamefont{Sch\"on}},\ and\ \bibinfo {author}
  {\bibfnamefont{J.}~\bibnamefont{von Delft}},\ }%
  \bibfield{journal}{%
  \Doi{10.1103/PhysRevLett.91.247202}{\bibinfo {journal} {Phys. Rev. Lett.}}\
  }%
  \textbf{\bibinfo {volume} {91}},\ \bibinfo {pages} {247202}\  (\bibinfo
  {year} {2003})%
  \bibAnnoteFile{NoStop}{martinek_kondo_2003-1}%
\bibitem{martinek_gate-controlled_2005}%
  \BibitemOpen
  \bibfield{author}{%
  \bibinfo {author} {\bibfnamefont{J.}~\bibnamefont{Martinek}}, \bibinfo
  {author} {\bibfnamefont{M.}~\bibnamefont{Sindel}}, \bibinfo {author}
  {\bibfnamefont{L.}~\bibnamefont{Borda}}, \bibinfo {author}
  {\bibfnamefont{J.}~\bibnamefont{Barnas}}, \bibinfo {author}
  {\bibfnamefont{R.}~\bibnamefont{Bulla}}, \bibinfo {author}
  {\bibfnamefont{J.}~\bibnamefont{K\"onig}}, \bibinfo {author}
  {\bibfnamefont{G.}~\bibnamefont{Sch\"on}}, \bibinfo {author}
  {\bibfnamefont{S.}~\bibnamefont{Maekawa}},\ and\ \bibinfo {author}
  {\bibfnamefont{J.}~\bibnamefont{von Delft}},\ }%
  \bibfield{journal}{%
  \Doi{10.1103/PhysRevB.72.121302}{\bibinfo {journal} {Phys. Rev. B.}}\ }%
  \textbf{\bibinfo {volume} {72}},\ \bibinfo {pages} {121302}\  (\bibinfo
  {year} {2005})%
  \bibAnnoteFile{NoStop}{martinek_gate-controlled_2005}%
\bibitem{utsumi_nonequilibrium_2005}%
  \BibitemOpen
  \bibfield{author}{%
  \bibinfo {author} {\bibfnamefont{Y.}~\bibnamefont{Utsumi}}, \bibinfo {author}
  {\bibfnamefont{J.}~\bibnamefont{Martinek}}, \bibinfo {author}
  {\bibfnamefont{G.}~\bibnamefont{Sch\"on}}, \bibinfo {author}
  {\bibfnamefont{H.}~\bibnamefont{Imamura}},\ and\ \bibinfo {author}
  {\bibfnamefont{S.}~\bibnamefont{Maekawa}},\ }%
  \bibfield{journal}{%
  \Doi{10.1103/PhysRevB.71.245116}{\bibinfo {journal} {Phys. Rev. B.}}\ }%
  \textbf{\bibinfo {volume} {71}},\ \bibinfo {pages} {245116}\  (\bibinfo
  {year} {2005})%
  \bibAnnoteFile{NoStop}{utsumi_nonequilibrium_2005}%
\bibitem{sindel_kondo_2007}%
  \BibitemOpen
  \bibfield{author}{%
  \bibinfo {author} {\bibfnamefont{M.}~\bibnamefont{Sindel}}, \bibinfo {author}
  {\bibfnamefont{L.}~\bibnamefont{Borda}}, \bibinfo {author}
  {\bibfnamefont{J.}~\bibnamefont{Martinek}}, \bibinfo {author}
  {\bibfnamefont{R.}~\bibnamefont{Bulla}}, \bibinfo {author}
  {\bibfnamefont{J.}~\bibnamefont{K\"onig}}, \bibinfo {author}
  {\bibfnamefont{G.}~\bibnamefont{Sch\"on}}, \bibinfo {author}
  {\bibfnamefont{S.}~\bibnamefont{Maekawa}},\ and\ \bibinfo {author}
  {\bibfnamefont{J.}~\bibnamefont{von Delft}},\ }%
  \bibfield{journal}{%
  \Doi{10.1103/PhysRevB.76.045321}{\bibinfo {journal} {Phys. Rev. B.}}\ }%
  \textbf{\bibinfo {volume} {76}},\ \bibinfo {pages} {045321}\  (\bibinfo
  {year} {2007})%
  \bibAnnoteFile{NoStop}{sindel_kondo_2007}%
\bibitem{fazio_resonant_1998}%
  \BibitemOpen
  \bibfield{author}{%
  \bibinfo {author} {\bibfnamefont{R.}~\bibnamefont{Fazio}}\ and\ \bibinfo
  {author} {\bibfnamefont{R.}~\bibnamefont{Raimondi}},\ }%
  \bibfield{journal}{%
  \Doi{10.1103/PhysRevLett.80.2913}{\bibinfo {journal} {Phys. Rev. Lett.}}\ }%
  \textbf{\bibinfo {volume} {80}},\ \bibinfo {pages} {2913}\  (\bibinfo {year}
  {1998})%
  \bibAnnoteFile{NoStop}{fazio_resonant_1998}%
\bibitem{fazio_erratum:_1999}%
  \BibitemOpen
  \bibfield{author}{%
  \bibinfo {author} {\bibfnamefont{R.}~\bibnamefont{Fazio}}\ and\ \bibinfo
  {author} {\bibfnamefont{R.}~\bibnamefont{Raimondi}},\ }%
  \bibfield{journal}{%
  \Doi{10.1103/PhysRevLett.82.4950}{\bibinfo {journal} {Phys. Rev. Lett.}}\ }%
  \textbf{\bibinfo {volume} {82}},\ \bibinfo {pages} {4950}\  (\bibinfo {year}
  {1999})%
  \bibAnnoteFile{NoStop}{fazio_erratum:_1999}%
\bibitem{kang_kondo_1998}%
  \BibitemOpen
  \bibfield{author}{%
  \bibinfo {author} {\bibfnamefont{K.}~\bibnamefont{Kang}},\ }%
  \bibfield{journal}{%
  \Doi{10.1103/PhysRevB.58.9641}{\bibinfo {journal} {Phys. Rev. B.}}\ }%
  \textbf{\bibinfo {volume} {58}},\ \bibinfo {pages} {9641}\  (\bibinfo {year}
  {1998})%
  \bibAnnoteFile{NoStop}{kang_kondo_1998}%
\bibitem{schwab_andreev_1999}%
  \BibitemOpen
  \bibfield{author}{%
  \bibinfo {author} {\bibfnamefont{P.}~\bibnamefont{Schwab}}\ and\ \bibinfo
  {author} {\bibfnamefont{R.}~\bibnamefont{Raimondi}},\ }%
  \bibfield{journal}{%
  \Doi{10.1103/PhysRevB.59.1637}{\bibinfo {journal} {Phys. Rev. B.}}\ }%
  \textbf{\bibinfo {volume} {59}},\ \bibinfo {pages} {1637}\  (\bibinfo {year}
  {1999})%
  \bibAnnoteFile{NoStop}{schwab_andreev_1999}%
\bibitem{clerk_andreev_2000}%
  \BibitemOpen
  \bibfield{author}{%
  \bibinfo {author} {\bibfnamefont{A.~A.}\ \bibnamefont{Clerk}}, \bibinfo
  {author} {\bibfnamefont{V.}~\bibnamefont{Ambegaokar}},\ and\ \bibinfo
  {author} {\bibfnamefont{S.}~\bibnamefont{Hershfield}},\ }%
  \bibfield{journal}{%
  \Doi{10.1103/PhysRevB.61.3555}{\bibinfo {journal} {Phys. Rev. B.}}\ }%
  \textbf{\bibinfo {volume} {61}},\ \bibinfo {pages} {3555}\  (\bibinfo {year}
  {2000})%
  \bibAnnoteFile{NoStop}{clerk_andreev_2000}%
\bibitem{cuevas_kondo_2001}%
  \BibitemOpen
  \bibfield{author}{%
  \bibinfo {author} {\bibfnamefont{J.~C.}\ \bibnamefont{Cuevas}}, \bibinfo
  {author} {\bibfnamefont{A.}\ \bibnamefont{Levy Yeyati}},\ and\ \bibinfo
  {author} {\bibfnamefont{A.}~\bibnamefont{{Martin-Rodero}}},\ }%
  \bibfield{journal}{%
  \Doi{10.1103/PhysRevB.63.094515}{\bibinfo {journal} {Phys. Rev. B.}}\ }%
  \textbf{\bibinfo {volume} {63}},\ \bibinfo {pages} {094515}\  (\bibinfo
  {year} {2001})%
  \bibAnnoteFile{NoStop}{cuevas_kondo_2001}%
\bibitem{pala_nonequilibrium_2007}%
  \BibitemOpen
  \bibfield{author}{%
  \bibinfo {author} {\bibfnamefont{M.~G.}\ \bibnamefont{Pala}}, \bibinfo
  {author} {\bibfnamefont{M.}~\bibnamefont{Governale}},\ and\ \bibinfo {author}
  {\bibfnamefont{J.}~\bibnamefont{K\"onig}},\ }%
  \bibfield{journal}{%
  \Doi{10.1088/1367-2630/9/8/278}{\bibinfo {journal} {New J. Phys.}}\ }%
  \textbf{\bibinfo {volume} {9}},\ \bibinfo {pages} {278}\  (\bibinfo {year}
  {2007})%
  \bibAnnoteFile{NoStop}{pala_nonequilibrium_2007}%
\bibitem{governale_real-time_2008}%
  \BibitemOpen
  \bibfield{author}{%
  \bibinfo {author} {\bibfnamefont{M.}~\bibnamefont{Governale}}, \bibinfo
  {author} {\bibfnamefont{M.~G.}\ \bibnamefont{Pala}},\ and\ \bibinfo {author}
  {\bibfnamefont{J.}~\bibnamefont{K\"onig}},\ }%
  \bibfield{journal}{%
  \Doi{10.1103/PhysRevB.77.134513}{\bibinfo {journal} {Phys. Rev. B.}}\ }%
  \textbf{\bibinfo {volume} {77}},\ \bibinfo {pages} {134513}\  (\bibinfo
  {year} {2008})%
  \bibAnnoteFile{NoStop}{governale_real-time_2008}%
\bibitem{futterer_nonlocal_2009}%
  \BibitemOpen
  \bibfield{author}{%
  \bibinfo {author} {\bibfnamefont{D.}~\bibnamefont{Futterer}}, \bibinfo
  {author} {\bibfnamefont{M.}~\bibnamefont{Governale}}, \bibinfo {author}
  {\bibfnamefont{M.~G.}\ \bibnamefont{Pala}},\ and\ \bibinfo {author}
  {\bibfnamefont{J.}~\bibnamefont{K\"onig}},\ }%
  \bibfield{journal}{%
  \Doi{10.1103/PhysRevB.79.054505}{\bibinfo {journal} {Phys. Rev. B.}}\ }%
  \textbf{\bibinfo {volume} {79}},\ \bibinfo {pages} {054505}\  (\bibinfo
  {year} {2009})%
  \bibAnnoteFile{NoStop}{futterer_nonlocal_2009}%
\bibitem{yeyati_resonant_1997}%
  \BibitemOpen
  \bibfield{author}{%
  \bibinfo {author} {\bibfnamefont{A.}\ \bibnamefont{Levy Yeyati}}, \bibinfo
  {author} {\bibfnamefont{J.~C.}\ \bibnamefont{Cuevas}}, \bibinfo {author}
  {\bibfnamefont{A.}~\bibnamefont{{Lopez-Davalos}}},\ and\ \bibinfo {author}
  {\bibfnamefont{A.}~\bibnamefont{{Martin-Rodero}}},\ }%
  \bibfield{journal}{%
  \Doi{10.1103/PhysRevB.55.R6137}{\bibinfo {journal} {Phys. Rev. B.}}\ }%
  \textbf{\bibinfo {volume} {55}},\ \bibinfo {pages} {R6137}\  (\bibinfo {year}
  {1997})%
  \bibAnnoteFile{NoStop}{yeyati_resonant_1997}%
\bibitem{johansson_resonant_1999}%
  \BibitemOpen
  \bibfield{author}{%
  \bibinfo {author} {\bibfnamefont{G.}~\bibnamefont{Johansson}}, \bibinfo
  {author} {\bibfnamefont{E.~N.}\ \bibnamefont{Bratus}}, \bibinfo {author}
  {\bibfnamefont{V.~S.}\ \bibnamefont{Shumeiko}},\ and\ \bibinfo {author}
  {\bibfnamefont{G.}~\bibnamefont{Wendin}},\ }%
  \bibfield{journal}{%
  \Doi{10.1103/PhysRevB.60.1382}{\bibinfo {journal} {Phys. Rev. B.}}\ }%
  \textbf{\bibinfo {volume} {60}},\ \bibinfo {pages} {1382}\  (\bibinfo {year}
  {1999})%
  \bibAnnoteFile{NoStop}{johansson_resonant_1999}%
\bibitem{clerk_loss_2000}%
  \BibitemOpen
  \bibfield{author}{%
  \bibinfo {author} {\bibfnamefont{A.~A.}\ \bibnamefont{Clerk}}\ and\ \bibinfo
  {author} {\bibfnamefont{V.}~\bibnamefont{Ambegaokar}},\ }%
  \bibfield{journal}{%
  \Doi{10.1103/PhysRevB.61.9109}{\bibinfo {journal} {Phys. Rev. B.}}\ }%
  \textbf{\bibinfo {volume} {61}},\ \bibinfo {pages} {9109}\  (\bibinfo {year}
  {2000})%
  \bibAnnoteFile{NoStop}{clerk_loss_2000}%
\bibitem{avishai_superconductor-quantum_2003}%
  \BibitemOpen
  \bibfield{author}{%
  \bibinfo {author} {\bibfnamefont{Y.}~\bibnamefont{Avishai}}, \bibinfo
  {author} {\bibfnamefont{A.}~\bibnamefont{Golub}},\ and\ \bibinfo {author}
  {\bibfnamefont{A.~D.}\ \bibnamefont{Zaikin}},\ }%
  \bibfield{journal}{%
  \Doi{10.1103/PhysRevB.67.041301}{\bibinfo {journal} {Phys. Rev. B.}}\ }%
  \textbf{\bibinfo {volume} {67}},\ \bibinfo {pages} {041301}\  (\bibinfo
  {year} {2003})%
  \bibAnnoteFile{NoStop}{avishai_superconductor-quantum_2003}%
\bibitem{sellier_pi_2005}%
  \BibitemOpen
  \bibfield{author}{%
  \bibinfo {author} {\bibfnamefont{G.}~\bibnamefont{Sellier}}, \bibinfo
  {author} {\bibfnamefont{T.}~\bibnamefont{Kopp}}, \bibinfo {author}
  {\bibfnamefont{J.}~\bibnamefont{Kroha}},\ and\ \bibinfo {author}
  {\bibfnamefont{Y.~S.}\ \bibnamefont{Barash}},\ }%
  \bibfield{journal}{%
  \Doi{10.1103/PhysRevB.72.174502}{\bibinfo {journal} {Phys. Rev. B.}}\ }%
  \textbf{\bibinfo {volume} {72}},\ \bibinfo {pages} {174502}\  (\bibinfo
  {year} {2005})%
  \bibAnnoteFile{NoStop}{sellier_pi_2005}%
\bibitem{nussinov_spin_2005}%
  \BibitemOpen
  \bibfield{author}{%
  \bibinfo {author} {\bibfnamefont{Z.}~\bibnamefont{Nussinov}}, \bibinfo
  {author} {\bibfnamefont{A.}~\bibnamefont{Shnirman}}, \bibinfo {author}
  {\bibfnamefont{D.~P.}\ \bibnamefont{Arovas}}, \bibinfo {author}
  {\bibfnamefont{A.~V.}\ \bibnamefont{Balatsky}},\ and\ \bibinfo {author}
  {\bibfnamefont{J.~X.}\ \bibnamefont{Zhu}},\ }%
  \bibfield{journal}{%
  \Doi{10.1103/PhysRevB.71.214520}{\bibinfo {journal} {Phys. Rev. B.}}\ }%
  \textbf{\bibinfo {volume} {71}},\ \bibinfo {pages} {214520}\  (\bibinfo
  {year} {2005})%
  \bibAnnoteFile{NoStop}{nussinov_spin_2005}%
\bibitem{bergeret_interplay_2006}%
  \BibitemOpen
  \bibfield{author}{%
  \bibinfo {author} {\bibfnamefont{F.~S.}\ \bibnamefont{Bergeret}}, \bibinfo
  {author} {\bibfnamefont{A.}\ \bibnamefont{Levy Yeyati}},\ and\ \bibinfo
  {author} {\bibfnamefont{A.}~\bibnamefont{{Martin-Rodero}}},\ }%
  \bibfield{journal}{%
  \Doi{10.1103/PhysRevB.74.132505}{\bibinfo {journal} {Phys. Rev. B.}}\ }%
  \textbf{\bibinfo {volume} {74}},\ \bibinfo {pages} {132505}\  (\bibinfo
  {year} {2006})%
  \bibAnnoteFile{NoStop}{bergeret_interplay_2006}%
\bibitem{lpez_josephson_2007}%
  \BibitemOpen
  \bibfield{author}{%
  \bibinfo {author} {\bibfnamefont{R.}~\bibnamefont{López}}, \bibinfo {author}
  {\bibfnamefont{M.~S.}~\bibnamefont{Choi}},\ and\ \bibinfo {author}
  {\bibfnamefont{R.}~\bibnamefont{Aguado}},\ }%
  \bibfield{journal}{%
  \Doi{10.1103/PhysRevB.75.045132}{\bibinfo {journal} {Phys. Rev. B.}}\ }%
  \textbf{\bibinfo {volume} {75}},\ \bibinfo {pages} {045132}\  (\bibinfo
  {year} {2007})%
  \bibAnnoteFile{NoStop}{lpez_josephson_2007}%
\bibitem{karrasch_josephson_2008}%
  \BibitemOpen
  \bibfield{author}{%
  \bibinfo {author} {\bibfnamefont{C.}~\bibnamefont{Karrasch}}, \bibinfo
  {author} {\bibfnamefont{A.}~\bibnamefont{Oguri}},\ and\ \bibinfo {author}
  {\bibfnamefont{V.}~\bibnamefont{Meden}},\ }%
  \bibfield{journal}{%
  \Doi{10.1103/PhysRevB.77.024517}{\bibinfo {journal} {Phys. Rev. B.}}\ }%
  \textbf{\bibinfo {volume} {77}},\ \bibinfo {pages} {024517}\  (\bibinfo
  {year} {2008})%
  \bibAnnoteFile{NoStop}{karrasch_josephson_2008}%
\bibitem{buitelaar_quantum_2002}%
  \BibitemOpen
  \bibfield{author}{%
  \bibinfo {author} {\bibfnamefont{M.~R.}\ \bibnamefont{Buitelaar}}, \bibinfo
  {author} {\bibfnamefont{T.}~\bibnamefont{Nussbaumer}},\ and\ \bibinfo
  {author} {\bibfnamefont{C.}~\bibnamefont{Sch\"onenberger}},\ }%
  \bibfield{journal}{%
  \Doi{10.1103/PhysRevLett.89.256801}{\bibinfo {journal} {Phys. Rev. Lett.}}\
  }%
  \textbf{\bibinfo {volume} {89}},\ \bibinfo {pages} {256801}\  (\bibinfo
  {year} {2002})%
  \bibAnnoteFile{NoStop}{buitelaar_quantum_2002}%
\bibitem{cleuziou_carbon_2006}%
  \BibitemOpen
  \bibfield{author}{%
  \bibinfo {author} {\bibfnamefont{J.}~\bibnamefont{Cleuziou}}, \bibinfo
  {author} {\bibfnamefont{W.}~\bibnamefont{Wernsdorfer}}, \bibinfo {author}
  {\bibfnamefont{V.}~\bibnamefont{Bouchiat}}, \bibinfo {author}
  {\bibfnamefont{T.}~\bibnamefont{Ondarcuhu}},\ and\ \bibinfo {author}
  {\bibfnamefont{M.}~\bibnamefont{Monthioux}},\ }%
  \bibfield{journal}{%
  \Doi{10.1038/nnano.2006.54}{\bibinfo {journal} {Nat. Nano.}}\ }%
  \textbf{\bibinfo {volume} {1}},\ \bibinfo {pages} {53}\  (\bibinfo {year}
  {2006})%
  \bibAnnoteFile{NoStop}{cleuziou_carbon_2006}%
\bibitem{jarillo-herrero_quantum_2006}%
  \BibitemOpen
  \bibfield{author}{%
  \bibinfo {author} {\bibfnamefont{P.}~\bibnamefont{{Jarillo-Herrero}}},
  \bibinfo {author} {\bibfnamefont{J.~A.}\ \bibnamefont{van Dam}},\ and\
  \bibinfo {author} {\bibfnamefont{L.~P.}\ \bibnamefont{Kouwenhoven}},\ }%
  \bibfield{journal}{%
  \Doi{10.1038/nature04550}{\bibinfo {journal} {Nature}}\ }%
  \textbf{\bibinfo {volume} {439}},\ \bibinfo {pages} {953}\  (\bibinfo {year}
  {2006})%
  \bibAnnoteFile{NoStop}{jarillo-herrero_quantum_2006}%
\bibitem{jrgensen_electron_2006}%
  \BibitemOpen
  \bibfield{author}{%
  \bibinfo {author} {\bibfnamefont{H.~I.}\ \bibnamefont{Jørgensen}}, \bibinfo
  {author} {\bibfnamefont{K.}~\bibnamefont{{Grove-Rasmussen}}}, \bibinfo
  {author} {\bibfnamefont{T.}~\bibnamefont{Novotny}}, \bibinfo {author}
  {\bibfnamefont{K.}~\bibnamefont{Flensberg}},\ and\ \bibinfo {author}
  {\bibfnamefont{P.~E.}\ \bibnamefont{Lindelof}},\ }%
  \bibfield{journal}{%
  \Doi{10.1103/PhysRevLett.96.207003}{\bibinfo {journal} {Phys. Rev. Lett.}}\
  }%
  \textbf{\bibinfo {volume} {96}},\ \bibinfo {pages} {207003}\  (\bibinfo
  {year} {2006})%
  \bibAnnoteFile{NoStop}{jrgensen_electron_2006}%
\bibitem{jrgensen_critical_2007}%
  \BibitemOpen
  \bibfield{author}{%
  \bibinfo {author} {\bibfnamefont{H.}\ \bibnamefont{Ingerslev Jørgensen}}, \bibinfo
  {author} {\bibfnamefont{T.}~\bibnamefont{Novotny}}, \bibinfo {author}
  {\bibfnamefont{K.}~\bibnamefont{{Grove-Rasmussen}}}, \bibinfo {author}
  {\bibfnamefont{K.}~\bibnamefont{Flensberg}},\ and\ \bibinfo {author}
  {\bibfnamefont{P.~E.}\ \bibnamefont{Lindelof}},\ }%
  \bibfield{journal}{%
  \Doi{10.1021/nl071152w}{\bibinfo {journal} {Nano Lett.}}\ }%
  \textbf{\bibinfo {volume} {7}},\ \bibinfo {pages} {2441}\  (\bibinfo {year}
  {2007})%
  \bibAnnoteFile{NoStop}{jrgensen_critical_2007}%
\bibitem{dirks_superconducting_2009}%
  \BibitemOpen
  \bibfield{author}{%
  \bibinfo {author} {\bibfnamefont{T.}~\bibnamefont{Dirks}}, \bibinfo {author}
  {\bibfnamefont{Y.}~\bibnamefont{Chen}}, \bibinfo {author}
  {\bibfnamefont{N.~O.}\ \bibnamefont{Birge}},\ and\ \bibinfo {author}
  {\bibfnamefont{N.}~\bibnamefont{Mason}},\ }%
  \bibfield{journal}{%
  \Doi{10.1063/1.3253705}{\bibinfo {journal} {Appl. Phys. Lett.}}\ }%
  \textbf{\bibinfo {volume} {95}},\ \bibinfo {pages} {192103}\  (\bibinfo
  {year} {2009})%
  \bibAnnoteFile{NoStop}{dirks_superconducting_2009}%
\bibitem{herrmann_carbon_2010}%
  \BibitemOpen
  \bibfield{author}{%
  \bibinfo {author} {\bibfnamefont{L.~G.}\ \bibnamefont{Herrmann}}, \bibinfo
  {author} {\bibfnamefont{F.}~\bibnamefont{Portier}}, \bibinfo {author}
  {\bibfnamefont{P.}~\bibnamefont{Roche}}, \bibinfo {author}
  {\bibfnamefont{A.}\ \bibnamefont{Levy Yeyati}}, \bibinfo {author}
  {\bibfnamefont{T.}~\bibnamefont{Kontos}},\ and\ \bibinfo {author}
  {\bibfnamefont{C.}~\bibnamefont{Strunk}},\ }%
  \bibfield{journal}{%
  \Doi{10.1103/PhysRevLett.104.026801}{\bibinfo {journal} {Phys. Rev. Lett.}}\
  }%
  \textbf{\bibinfo {volume} {104}},\ \bibinfo {pages} {026801}\  (\bibinfo
  {year} {2010})%
  \bibAnnoteFile{NoStop}{herrmann_carbon_2010}%
\bibitem{dirks_andreev_2010}%
  \BibitemOpen
  \bibfield{author}{%
  \bibinfo {author} {\bibfnamefont{T.}~\bibnamefont{Dirks}}, \bibinfo {author}
  {\bibfnamefont{T.~L.}\ \bibnamefont{Hughes}}, \bibinfo {author}
  {\bibfnamefont{S.}~\bibnamefont{Lal}}, \bibinfo {author}
  {\bibfnamefont{B.}~\bibnamefont{Uchoa}}, \bibinfo {author}
  {\bibfnamefont{Y.}~\bibnamefont{Chen}}, \bibinfo {author}
  {\bibfnamefont{C.}~\bibnamefont{Chialvo}}, \bibinfo {author}
  {\bibfnamefont{P.~M.}\ \bibnamefont{Goldbart}},\ and\ \bibinfo {author}
  {\bibfnamefont{N.}~\bibnamefont{Mason}},\ }%
  \bibfield{journal}{%
  \bibinfo {journal} {1005.2749}}%
  \  (\bibinfo {year} {2010})%
  \bibAnnoteFile{NoStop}{dirks_andreev_2010}%
\bibitem{van_dam_supercurrent_2006}%
  \BibitemOpen
  \bibfield{author}{%
  \bibinfo {author} {\bibfnamefont{J.~A.}\ \bibnamefont{van Dam}}, \bibinfo
  {author} {\bibfnamefont{Y.~V.}\ \bibnamefont{Nazarov}}, \bibinfo {author}
  {\bibfnamefont{E.~P. A.~M.}\ \bibnamefont{Bakkers}}, \bibinfo {author}
  {\bibfnamefont{S.~D.}\ \bibnamefont{Franceschi}},\ and\ \bibinfo {author}
  {\bibfnamefont{L.~P.}\ \bibnamefont{Kouwenhoven}},\ }%
  \bibfield{journal}{%
  \Doi{10.1038/nature05018}{\bibinfo {journal} {Nature}}\ }%
  \textbf{\bibinfo {volume} {442}},\ \bibinfo {pages} {667}\  (\bibinfo {year}
  {2006})%
  \bibAnnoteFile{NoStop}{van_dam_supercurrent_2006}%
\bibitem{sand-jespersen_kondo-enhanced_2007}%
  \BibitemOpen
  \bibfield{author}{%
  \bibinfo {author} {\bibfnamefont{T.}~\bibnamefont{{Sand-Jespersen}}},
  \bibinfo {author} {\bibfnamefont{J.}~\bibnamefont{Paaske}}, \bibinfo {author}
  {\bibfnamefont{B.~M.}\ \bibnamefont{Andersen}}, \bibinfo {author}
  {\bibfnamefont{K.}~\bibnamefont{{Grove-Rasmussen}}}, \bibinfo {author}
  {\bibfnamefont{H.~I.}\ \bibnamefont{Jorgensen}}, \bibinfo {author}
  {\bibfnamefont{M.}~\bibnamefont{Aagesen}}, \bibinfo {author}
  {\bibfnamefont{C.~B.}\ \bibnamefont{Sorensen}}, \bibinfo {author}
  {\bibfnamefont{P.~E.}\ \bibnamefont{Lindelof}}, \bibinfo {author}
  {\bibfnamefont{K.}~\bibnamefont{Flensberg}},\ and\ \bibinfo {author}
  {\bibfnamefont{J.}~\bibnamefont{Nygard}},\ }%
  \bibfield{journal}{%
  \Doi{10.1103/PhysRevLett.99.126603}{\bibinfo {journal} {Phys. Rev. Lett.}}\
  }%
  \textbf{\bibinfo {volume} {99}},\ \bibinfo {pages} {126603}\  (\bibinfo
  {year} {2007})%
  \bibAnnoteFile{NoStop}{sand-jespersen_kondo-enhanced_2007}%
\bibitem{buizert_kondo_2007}%
  \BibitemOpen
  \bibfield{author}{%
  \bibinfo {author} {\bibfnamefont{C.}~\bibnamefont{Buizert}}, \bibinfo
  {author} {\bibfnamefont{A.}~\bibnamefont{Oiwa}}, \bibinfo {author}
  {\bibfnamefont{K.}~\bibnamefont{Shibata}}, \bibinfo {author}
  {\bibfnamefont{K.}~\bibnamefont{Hirakawa}},\ and\ \bibinfo {author}
  {\bibfnamefont{S.}~\bibnamefont{Tarucha}},\ }%
  \bibfield{journal}{%
  \Doi{10.1103/PhysRevLett.99.136806}{\bibinfo {journal} {Phys. Rev. Lett.}}\
  }%
  \textbf{\bibinfo {volume} {99}},\ \bibinfo {pages} {136806}\  (\bibinfo
  {year} {2007})%
  \bibAnnoteFile{NoStop}{buizert_kondo_2007}%
\bibitem{winkelmann_superconductivity_2009}%
  \BibitemOpen
  \bibfield{author}{%
  \bibinfo {author} {\bibfnamefont{C.~B.}\ \bibnamefont{Winkelmann}}, \bibinfo
  {author} {\bibfnamefont{N.}~\bibnamefont{Roch}}, \bibinfo {author}
  {\bibfnamefont{W.}~\bibnamefont{Wernsdorfer}}, \bibinfo {author}
  {\bibfnamefont{V.}~\bibnamefont{Bouchiat}},\ and\ \bibinfo {author}
  {\bibfnamefont{F.}~\bibnamefont{Balestro}},\ }%
  \bibfield{journal}{%
  \Doi{10.1038/nphys1433}{\bibinfo {journal} {Nat. Phys.}}\ }%
  \textbf{\bibinfo {volume} {5}},\ \bibinfo {pages} {876}\  (\bibinfo {year}
  {2009})%
  \bibAnnoteFile{NoStop}{winkelmann_superconductivity_2009}%
\bibitem{knig_zero-bias_1996}%
  \BibitemOpen
  \bibfield{author}{%
  \bibinfo {author} {\bibfnamefont{J.}~\bibnamefont{K\"onig}}, \bibinfo {author}
  {\bibfnamefont{H.}~\bibnamefont{Schoeller}},\ and\ \bibinfo {author}
  {\bibfnamefont{G.}~\bibnamefont{Sch\"on}},\ }%
  \bibfield{journal}{%
  \Doi{10.1103/PhysRevLett.76.1715}{\bibinfo {journal} {Phys. Rev. Lett.}}\ }%
  \textbf{\bibinfo {volume} {76}},\ \bibinfo {pages} {1715}\  (\bibinfo {year}
  {1996})%
  \bibAnnoteFile{NoStop}{knig_zero-bias_1996}%
\bibitem{knig_resonant_1996}%
  \BibitemOpen
  \bibfield{author}{%
  \bibinfo {author} {\bibfnamefont{J.}~\bibnamefont{K\"onig}}, \bibinfo {author}
  {\bibfnamefont{J.}~\bibnamefont{Schmid}}, \bibinfo {author}
  {\bibfnamefont{H.}~\bibnamefont{Schoeller}},\ and\ \bibinfo {author}
  {\bibfnamefont{G.}~\bibnamefont{Sch\"on}},\ }%
  \bibfield{journal}{%
  \Doi{10.1103/PhysRevB.54.16820}{\bibinfo {journal} {Phys. Rev. B.}}\ }%
  \textbf{\bibinfo {volume} {54}},\ \bibinfo {pages} {16820}\  (\bibinfo {year}
  {1996})%
  \bibAnnoteFile{NoStop}{knig_resonant_1996}%
\bibitem{schoeller_transport_1997}%
  \BibitemOpen
  \bibfield{author}{%
  \bibinfo {author} {\bibfnamefont{H.}~\bibnamefont{Schoeller}},\ }%
  \emph{\bibinfo {title} {Transport theory of interacting quantum dots}}\
  (\bibinfo {year} {1997})%
  \bibAnnoteFile{NoStop}{schoeller_transport_1997}%
\bibitem{knig_quantum_1999}%
  \BibitemOpen
  \bibfield{author}{%
  \bibinfo {author} {\bibfnamefont{J.}~\bibnamefont{K\"onig}},\ }%
  \emph{\bibinfo {title} {Quantum Fluctuations in the {Single-Electron}
  Transistor}}\ (\bibinfo {publisher} {Shaker},\ \bibinfo {address} {Aachen},\
  \bibinfo {year} {1999})%
  \bibAnnoteFile{NoStop}{knig_quantum_1999}%
\bibitem{rozhkov_interacting-impurity_2000}%
  \BibitemOpen
  \bibfield{author}{%
  \bibinfo {author} {\bibfnamefont{A.~V.}\ \bibnamefont{Rozhkov}}\ and\
  \bibinfo {author} {\bibfnamefont{D.~P.}\ \bibnamefont{Arovas}},\ }%
  \bibfield{journal}{%
  \Doi{10.1103/PhysRevB.62.6687}{\bibinfo {journal} {Phys. Rev. B.}}\ }%
  \textbf{\bibinfo {volume} {62}},\ \bibinfo {pages} {6687}\  (\bibinfo {year}
  {2000})%
  \bibAnnoteFile{NoStop}{rozhkov_interacting-impurity_2000}%
\bibitem{karrasch_supercurrent_2009}%
  \BibitemOpen
  \bibfield{author}{%
  \bibinfo {author} {\bibfnamefont{C.}~\bibnamefont{Karrasch}}\ and\ \bibinfo
  {author} {\bibfnamefont{V.}~\bibnamefont{Meden}},\ }%
  \bibfield{journal}{%
  \Doi{10.1103/PhysRevB.79.045110}{\bibinfo {journal} {Phys. Rev. B.}}\ }%
  \textbf{\bibinfo {volume} {79}},\ \bibinfo {pages} {045110}\  (\bibinfo
  {year} {2009})%
  \bibAnnoteFile{NoStop}{karrasch_supercurrent_2009}%
\bibitem{meng_self-consistent_2009}%
  \BibitemOpen
  \bibfield{author}{%
  \bibinfo {author} {\bibfnamefont{T.}~\bibnamefont{Meng}}, \bibinfo {author}
  {\bibfnamefont{S.}~\bibnamefont{Florens}},\ and\ \bibinfo {author}
  {\bibfnamefont{P.}~\bibnamefont{Simon}},\ }%
  \bibfield{journal}{%
  \Doi{10.1103/PhysRevB.79.224521}{\bibinfo {journal} {Phys. Rev. B.}}\ }%
  \textbf{\bibinfo {volume} {79}},\ \bibinfo {pages} {224521}\  (\bibinfo
  {year} {2009})%
  \bibAnnoteFile{NoStop}{meng_self-consistent_2009}%
\bibitem{erlingsson_hyperfine-mediated_2002}%
  \BibitemOpen
  \bibfield{author}{%
  \bibinfo {author} {\bibfnamefont{S.~I.}\ \bibnamefont{Erlingsson}}\ and\
  \bibinfo {author} {\bibfnamefont{Y.~V.}\ \bibnamefont{Nazarov}},\ }%
  \bibfield{journal}{%
  \Doi{10.1103/PhysRevB.66.155327}{\bibinfo {journal} {Phys. Rev. B.}}\ }%
  \textbf{\bibinfo {volume} {66}},\ \bibinfo {pages} {155327}\  (\bibinfo
  {year} {2002})%
  \bibAnnoteFile{NoStop}{erlingsson_hyperfine-mediated_2002}%
\bibitem{merkulov_electron_2002}%
  \BibitemOpen
  \bibfield{author}{%
  \bibinfo {author} {\bibfnamefont{I.~A.}\ \bibnamefont{Merkulov}}, \bibinfo
  {author} {\bibfnamefont{A.~L.}\ \bibnamefont{Efros}},\ and\ \bibinfo {author}
  {\bibfnamefont{M.}~\bibnamefont{Rosen}},\ }%
  \bibfield{journal}{%
  \Doi{10.1103/PhysRevB.65.205309}{\bibinfo {journal} {Phys. Rev. B.}}\ }%
  \textbf{\bibinfo {volume} {65}},\ \bibinfo {pages} {205309}\  (\bibinfo
  {year} {2002})%
  \bibAnnoteFile{NoStop}{merkulov_electron_2002}%
\bibitem{khaetskii_electron_2003}%
  \BibitemOpen
  \bibfield{author}{%
  \bibinfo {author} {\bibfnamefont{A.}~\bibnamefont{Khaetskii}}, \bibinfo
  {author} {\bibfnamefont{D.}~\bibnamefont{Loss}},\ and\ \bibinfo {author}
  {\bibfnamefont{L.}~\bibnamefont{Glazman}},\ }%
  \bibfield{journal}{%
  \Doi{10.1103/PhysRevB.67.195329}{\bibinfo {journal} {Phys. Rev. B.}}\ }%
  \textbf{\bibinfo {volume} {67}},\ \bibinfo {pages} {195329}\  (\bibinfo
  {year} {2003})%
  \bibAnnoteFile{NoStop}{khaetskii_electron_2003}%
\bibitem{khaetskii_spin-flip_2001}%
  \BibitemOpen
  \bibfield{author}{%
  \bibinfo {author} {\bibfnamefont{A.~V.}\ \bibnamefont{Khaetskii}}\ and\
  \bibinfo {author} {\bibfnamefont{Y.~V.}\ \bibnamefont{Nazarov}},\ }%
  \bibfield{journal}{%
  \Doi{10.1103/PhysRevB.64.125316}{\bibinfo {journal} {Phys. Rev. B.}}\ }%
  \textbf{\bibinfo {volume} {64}},\ \bibinfo {pages} {125316}\  (\bibinfo
  {year} {2001})%
  \bibAnnoteFile{NoStop}{khaetskii_spin-flip_2001}%
\bibitem{golovach_phonon-induced_2004}%
  \BibitemOpen
  \bibfield{author}{%
  \bibinfo {author} {\bibfnamefont{V.~N.}\ \bibnamefont{Golovach}}, \bibinfo
  {author} {\bibfnamefont{A.}~\bibnamefont{Khaetskii}},\ and\ \bibinfo {author}
  {\bibfnamefont{D.}~\bibnamefont{Loss}},\ }%
  \bibfield{journal}{%
  \Doi{10.1103/PhysRevLett.93.016601}{\bibinfo {journal} {Phys. Rev. Lett.}}\
  }%
  \textbf{\bibinfo {volume} {93}},\ \bibinfo {pages} {016601}\  (\bibinfo
  {year} {2004})%
  \bibAnnoteFile{NoStop}{golovach_phonon-induced_2004}%
\end{thebibliography}
\end{document}